\documentclass[preprint2]{aastex62}

\usepackage{amsmath,bm,multirow,soul}
\usepackage{booktabs}
\usepackage[flushleft]{threeparttable}
\usepackage{url}
\usepackage{hyperref}
\RequirePackage{lineno}
\accepted{07 Jan 2019}

\shorttitle{1ES\,1215$+$303 Long-term study}
\shortauthors{VERITAS, {\it{Fermi}}-LAT et al.}

\begin{document}

\title{A decade of multi-wavelength observations of the TeV blazar 1ES\,1215+303: Extreme shift of the synchrotron peak frequency and long-term optical-gamma-ray flux increase}

\correspondingauthor{Janeth Valverde, Qi Feng, Olivier Hervet}
\email{valverde@llr.in2p3.fr, qifeng@nevis.columbia.edu, ohervet@ucsc.edu}

\author{Janeth Valverde}
\affiliation{Laboratoire Leprince-Ringuet, \'Ecole Polytechnique, CNRS/IN2P3, 91128 Palaiseau, France}

\author{Deirdre Horan}
\affiliation{Laboratoire Leprince-Ringuet, \'Ecole Polytechnique, CNRS/IN2P3, 91128 Palaiseau, France}

\author{Denis Bernard}
\affiliation{Laboratoire Leprince-Ringuet, \'Ecole Polytechnique, CNRS/IN2P3, 91128 Palaiseau, France}

\author{Stephen Fegan}
\affiliation{Laboratoire Leprince-Ringuet, \'Ecole Polytechnique, CNRS/IN2P3, 91128 Palaiseau, France}

\collaboration{({\it{Fermi}}-LAT collaboration)}

\author{A.~U.~Abeysekara}\affiliation{Department of Physics and Astronomy, University of Utah, Salt Lake City, UT 84112, USA}
\author{A.~Archer}\affiliation{Department of Physics and Astronomy, DePauw University, Greencastle, IN 46135-0037, USA}
\author{W.~Benbow}\affiliation{Center for Astrophysics $|$ Harvard \& Smithsonian, Cambridge, MA 02138, USA}
\author{R.~Bird}\affiliation{Department of Physics and Astronomy, University of California, Los Angeles, CA 90095, USA}
\author{A.~Brill}\affiliation{Physics Department, Columbia University, New York, NY 10027, USA}
\author{R.~Brose}\affiliation{Institute of Physics and Astronomy, University of Potsdam, 14476 Potsdam-Golm, Germany and DESY, Platanenallee 6, 15738 Zeuthen, Germany}
\author{M.~Buchovecky}\affiliation{Department of Physics and Astronomy, University of California, Los Angeles, CA 90095, USA}
\author{J.~H.~Buckley}\affiliation{Department of Physics, Washington University, St. Louis, MO 63130, USA}
\author{J.~L.~Christiansen}\affiliation{Physics Department, California Polytechnic State University, San Luis Obispo, CA 94307, USA}
\author{W.~Cui}\affiliation{Department of Physics and Astronomy, Purdue University, West Lafayette, IN 47907, USA and Department of Physics and Center for Astrophysics, Tsinghua University, Beijing 100084, China.}
\author{A.~Falcone}\affiliation{Department of Astronomy and Astrophysics, 525 Davey Lab, Pennsylvania State University, University Park, PA 16802, USA}
\author{Q.~Feng}\affiliation{Physics Department, Columbia University, New York, NY 10027, USA}
\author{J.~P.~Finley}\affiliation{Department of Physics and Astronomy, Purdue University, West Lafayette, IN 47907, USA}
\author{L.~Fortson}\affiliation{School of Physics and Astronomy, University of Minnesota, Minneapolis, MN 55455, USA}
\author{A.~Furniss}\affiliation{Department of Physics, California State University - East Bay, Hayward, CA 94542, USA}
\author{A.~Gent}\affiliation{School of Physics and Center for Relativistic Astrophysics, Georgia Institute of Technology, 837 State Street NW, Atlanta, GA 30332-0430}
\author{G.~H.~Gillanders}\affiliation{School of Physics, National University of Ireland Galway, University Road, Galway, Ireland}
\author{C.~Giuri}\affiliation{DESY, Platanenallee 6, 15738 Zeuthen, Germany}
\author{O.~Gueta}\affiliation{DESY, Platanenallee 6, 15738 Zeuthen, Germany}
\author{D.~Hanna}\affiliation{Physics Department, McGill University, Montreal, QC H3A 2T8, Canada}
\author{T.~Hassan}\affiliation{DESY, Platanenallee 6, 15738 Zeuthen, Germany}
\author{O.~Hervet}\affiliation{Santa Cruz Institute for Particle Physics and Department of Physics, University of California, Santa Cruz, CA 95064, USA}
\author{J.~Holder}\affiliation{Department of Physics and Astronomy and the Bartol Research Institute, University of Delaware, Newark, DE 19716, USA}
\author{G.~Hughes}\affiliation{Center for Astrophysics $|$ Harvard \& Smithsonian, Cambridge, MA 02138, USA}
\author{T.~B.~Humensky}\affiliation{Physics Department, Columbia University, New York, NY 10027, USA}
\author{P.~Kaaret}\affiliation{Department of Physics and Astronomy, University of Iowa, Van Allen Hall, Iowa City, IA 52242, USA}
\author{N.~Kelley-Hoskins}\affiliation{DESY, Platanenallee 6, 15738 Zeuthen, Germany}
\author{M.~Kertzman}\affiliation{Department of Physics and Astronomy, DePauw University, Greencastle, IN 46135-0037, USA}
\author{D.~Kieda}\affiliation{Department of Physics and Astronomy, University of Utah, Salt Lake City, UT 84112, USA}
\author{M.~Krause}\affiliation{DESY, Platanenallee 6, 15738 Zeuthen, Germany}
\author{F.~Krennrich}\affiliation{Department of Physics and Astronomy, Iowa State University, Ames, IA 50011, USA}
\author{M.~J.~Lang}\affiliation{School of Physics, National University of Ireland Galway, University Road, Galway, Ireland}
\author{G.~Maier}\affiliation{DESY, Platanenallee 6, 15738 Zeuthen, Germany}
\author{P.~Moriarty}\affiliation{School of Physics, National University of Ireland Galway, University Road, Galway, Ireland}
\author{R.~Mukherjee}\affiliation{Department of Physics and Astronomy, Barnard College, Columbia University, New York, NY 10027, USA}

\author{D.~Nieto}\affiliation{Institute of Particle and Cosmos Physics, Universidad Complutense de Madrid, 28040 Madrid, Spain}
\author{M.~Nievas-Rosillo}\affiliation{DESY, Platanenallee 6, 15738 Zeuthen, Germany}
\author{S.~O'Brien}\affiliation{Physics Department, McGill University, Montreal, QC H3A 2T8, Canada}
\author{R.~A.~Ong}\affiliation{Department of Physics and Astronomy, University of California, Los Angeles, CA 90095, USA}
\author{A.~N.~Otte}\affiliation{School of Physics and Center for Relativistic Astrophysics, Georgia Institute of Technology, 837 State Street NW, Atlanta, GA 30332-0430}
\author{N.~Park}\affiliation{WIPAC and Department of Physics, University of Wisconsin-Madison, Madison WI, USA}
\author{A.~Petrashyk}\affiliation{Physics Department, Columbia University, New York, NY 10027, USA}
\author{K.~Pfrang}\affiliation{DESY, Platanenallee 6, 15738 Zeuthen, Germany}
\author{A.~Pichel}\affiliation{Instituto de Astronomía y Física del Espacio (IAFE, CONICET-UBA), CC 67 - Suc. 28, (C1428ZAA) Ciudad Autónoma de Buenos Aires, Argentina}
\author{M.~Pohl}\affiliation{Institute of Physics and Astronomy, University of Potsdam, 14476 Potsdam-Golm, Germany and DESY, Platanenallee 6, 15738 Zeuthen, Germany}
\author{R.~R.~Prado}\affiliation{DESY, Platanenallee 6, 15738 Zeuthen, Germany}
\author{E.~Pueschel}\affiliation{DESY, Platanenallee 6, 15738 Zeuthen, Germany}
\author{J.~Quinn}\affiliation{School of Physics, University College Dublin, Belfield, Dublin 4, Ireland}
\author{K.~Ragan}\affiliation{Physics Department, McGill University, Montreal, QC H3A 2T8, Canada}
\author{P.~T.~Reynolds}\affiliation{Department of Physical Sciences, Cork Institute of Technology, Bishopstown, Cork, Ireland}
\author{D.~Ribeiro}\affiliation{Physics Department, Columbia University, New York, NY 10027, USA}
\author{G.~T.~Richards}\affiliation{Department of Physics and Astronomy and the Bartol Research Institute, University of Delaware, Newark, DE 19716, USA}
\author{E.~Roache}\affiliation{Center for Astrophysics $|$ Harvard \& Smithsonian, Cambridge, MA 02138, USA}
\author{I.~Sadeh}\affiliation{DESY, Platanenallee 6, 15738 Zeuthen, Germany}
\author{M.~Santander}\affiliation{Department of Physics and Astronomy, University of Alabama, Tuscaloosa, AL 35487, USA}
\author{S.~S.~Scott}\affiliation{Santa Cruz Institute for Particle Physics and Department of Physics, University of California, Santa Cruz, CA 95064, USA}
\author{G.~H.~Sembroski}\affiliation{Department of Physics and Astronomy, Purdue University, West Lafayette, IN 47907, USA}
\author{K.~Shahinyan}\affiliation{School of Physics and Astronomy, University of Minnesota, Minneapolis, MN 55455, USA}
\author{R.~Shang}\affiliation{Department of Physics and Astronomy, University of California, Los Angeles, CA 90095, USA}
\author{I.~Sushch}\affiliation{Institute of Physics and Astronomy, University of Potsdam, 14476 Potsdam-Golm, Germany}
\author{V.~V.~Vassiliev}\affiliation{Department of Physics and Astronomy, University of California, Los Angeles, CA 90095, USA}
\author{A.~Weinstein}\affiliation{Department of Physics and Astronomy, Iowa State University, Ames, IA 50011, USA}
\author{R.~M.~Wells}\affiliation{Department of Physics and Astronomy, Iowa State University, Ames, IA 50011, USA}
\author{P.~Wilcox}\affiliation{School of Physics and Astronomy, University of Minnesota, Minneapolis, MN 55455, USA}
\author{A.~Wilhelm}\affiliation{Institute of Physics and Astronomy, University of Potsdam, 14476 Potsdam-Golm, Germany and DESY, Platanenallee 6, 15738 Zeuthen, Germany}
\author{D.~A.~Williams}\affiliation{Santa Cruz Institute for Particle Physics and Department of Physics, University of California, Santa Cruz, CA 95064, USA}
\author{T.~J~Williamson}\affiliation{Department of Physics and Astronomy and the Bartol Research Institute, University of Delaware, Newark, DE 19716, USA}

\collaboration{(VERITAS collaboration)}

\author{Giuliana Noto}
\affiliation{Department of Physics and Astronomy, Barnard College, Columbia University, New York, NY 10027, USA}

\author{P.~G.~Edwards} 
\affiliation{CSIRO Astronomy and Space Science, Australia Telescope National Facility, P.O. Box 76, Epping, NSW 1710, Australia}

\author{B.~G.~Piner}
\affiliation{Department of Physics and Astronomy, Whittier College, 13406 E. Philadelphia Street, Whittier, CA 90608, USA}
\affiliation{Jet Propulsion Laboratory, California Institute of Technology, 4800 Oak Grove Drive, Pasadena, CA 91106, USA}

\author{V. Fallah Ramazani}
\affiliation{Finnish Centre for Astronomy with ESO (FINCA), University of Turku, Finland}

\author{T. Hovatta}
\affiliation{Finnish Centre for Astronomy with ESO (FINCA), University of Turku, Finland}
\affiliation{Aalto University Mets\"ahovi Radio Observatory, Mets\"ahovintie 114, FI-02540 Kylm\"al\"a, Finland}

\author{J. Jormanainen}
\affiliation{Tuorla Observatory, Department of Physics and Astronomy, University of Turku, Finland}
\affiliation{Finnish Centre for Astronomy with ESO (FINCA), University of Turku, Finland}

\author{E. Lindfors}
\affiliation{Finnish Centre for Astronomy with ESO (FINCA), University of Turku, Finland}

\author{K. Nilsson} 
\affiliation{Finnish Centre for Astronomy with ESO (FINCA), University of Turku, Finland}

\author{L. Takalo}
\affiliation{Tuorla Observatory, Department of Physics and Astronomy, University of Turku, Finland}

\author{Y.~Y.~Kovalev} 
\affiliation{Astro Space Center of Lebedev Physical Institute, Profsoyuznaya 84/32, 117997 Moscow, Russia }
\affiliation{Moscow Institute of Physics and Technology, Dolgoprudny, Institutsky per., 9, Moscow region, 141700, Russia}
\affiliation{Max-Planck-Institut f\"ur Radioastronomie, Auf dem H\"ugel 69, D-53121 Bonn, Germany }

\author{M.~L.~Lister} 
\affiliation{Purdue University, 525 Northwestern Avenue, West Lafayette, IN 47907, USA }

\author{A.~B.~Pushkarev} 
\affiliation{Crimean Astrophysical Observatory, 98409 Nauchny, Crimea, Russia }
\affiliation{Astro Space Center of Lebedev Physical Institute, Profsoyuznaya 84/32, 117997 Moscow, Russia }

\author{T.~Savolainen} 
\affiliation{Aalto University Mets\"ahovi Radio Observatory, Mets\"ahovintie 114, FI-02540 Kylm\"al\"a, Finland}
\affiliation{Aalto University Department of Electronics and Nanoengineering, PL 15500, FI-00076 Aalto, Finland }\affiliation{Max-Planck-Institut f\"ur Radioastronomie, Auf dem H\"ugel 69, D-53121 Bonn, Germany }

\author{S. Kiehlmann}
\affiliation{Institute of Astrophysics, Foundation for Research and Technology-Hellas, GR-71110 Heraklion, Greece}
\affiliation{Department of Physics, Univ. of Crete, GR-70013 Heraklion, Greece}
\affiliation{Owens Valley Radio Observatory, California Institute of Technology, Pasadena, CA 91125, USA}

\author{W. Max-Moerbeck}
\affiliation{Departamento de Astronom\'ia, Universidad de Chile, Camino El Observatorio 1515, Las Condes, Santiago, Chile}

\author{A. C. S. Readhead}
\affiliation{Owens Valley Radio Observatory, California Institute of Technology, Pasadena, CA 91125, USA}

\author{A. L\"ahteenm\"aki}
\affiliation{Aalto University Mets\"ahovi Radio Observatory, Mets\"ahovintie 114, FI-02540 Kylm\"al\"a, Finland}
\affiliation{Aalto University Department of Electronics and Nanoengineering, PL 15500, FI-00076 Aalto, Finland }

\author{M. Tornikoski}
\affiliation{Aalto University Mets\"ahovi Radio Observatory, Mets\"ahovintie 114, FI-02540 Kylm\"al\"a, Finland}

\begin{abstract}

Blazars are known for their variability on a wide range of timescales at all wavelengths. 
Most studies of TeV gamma-ray blazars focus on short timescales, especially during flares. 
With a decade of observations from the {\sl Fermi}-LAT and VERITAS, we present an extensive study of the long-term multi-wavelength radio-to-gamma-ray flux-density variability, with the addition of a couple of short-time radio-structure and optical polarization observations of the blazar 1ES\,1215$+$303 ($z=0.130$), with a focus on its gamma-ray emission from 100~MeV to 30~TeV. 
Multiple strong GeV gamma-ray flares, a long-term increase in the gamma-ray and optical flux baseline and a linear correlation between these two bands are observed over the ten-year period.
Typical HBL behaviors are identified in the radio morphology and broadband spectrum of the source. 
Three stationary features in the innermost jet are resolved by VLBA at 43.1, 22.2, and 15.3\,GHz. 
We employ a two-component synchrotron self-Compton model to describe different flux states of the source, including the epoch during which an extreme shift in energy of the synchrotron peak frequency from infrared to soft X-rays is observed. 

\end{abstract}


\section{Introduction} \label{sec:intro}

1ES\,1215+303 \citep[R.A. $= \rm 12^h17^m52.0819^s$, Dec. $= +30^o07^\prime 00^{\prime\prime}635$, J2000;][]{2011AJ....142...89P}, also known by many other names including Ton\,605, ON\,325, B2\,1215+30 and S3\,1215+30, is a blazar detected in the very-high-energy (VHE; $\gtrsim$100~GeV) gamma-ray band. 
Blazars, of which there are, at the time of writing, 72 known to emit VHE radiation\footnote{\url{http://tevcat.uchicago.edu}}, are the most numerous sources detected at these energies comprising approximately one third of the VHE sources. 1ES\,1215+303 was first discovered at VHE by MAGIC \citep[the Major Atmospheric Gamma Imaging Cherenkov;][]{Aleksic12} and has been monitored by the Very Energetic Radiation Imaging Telescope Array (VERITAS) at TeV energies since 2008.

The source exhibited one of the most luminous and large-amplitude flares $E \gtrsim 90$~GeV ever detected from a VHE blazar measured by VERITAS, when, in 2014, the TeV flux reached 2.4 times the Crab Nebula flux with a variability timescale of $< 3.6$\,h  \citep{Abeysekara17B21215}. 
In the high-energy (HE; $\approx\,$100\,MeV\,$-$\,$\approx\,$500\,GeV) gamma-ray band,  1ES\,1215+303 has been detected by the {\sl Fermi} Large Area Telescope (LAT), most recently as  4FGL\,J1217.9+3007  \citep{2019arXiv190210045T}. 
A high-flux state correlated with that detected in the VHE band was observed at these energies during the luminous and isolated gamma-ray flare of 2014 \citep{Abeysekara17B21215}. 

1ES\,1215+303 exhibits a double-humped spectral energy distribution (SED) typical of blazars, with the synchrotron peak between radio and X-ray energies and the high-energy peak at GeV\,$-$\,TeV energies. 
The synchrotron peak frequency of 1ES\,1215+303 has been measured to be $\nu_{\text{syn}}>10^{15}$ Hz which led to its classification as either an intermediate-frequency-peaked BL Lac\footnote{In the classification scheme of \cite{Padovani95}.} \citep[IBL; $\nu_{\text{syn}}=10^{15.58}$ Hz;][]{Nieppola06} or a high-synchrotron-peaked BL Lac\footnote{Classification based on the position of the synchrotron peak.} \citep[HBL; $\nu_{\text{syn}}=10^{15.205}$ Hz;][]{3LACAckermann15}. 
The redshift was measured to be $z = 0.13$ \citep{Akiyama03}, which was confirmed recently with high signal-to-noise ratio optical spectroscopic data \citep{Paiano17}, and from Ly$\alpha$ emission line at $z = 0.1305 \pm 0.0030$ \citep{Furniss2019}.\\

\begin{deluxetable*}{l|ccccc}
\tablecaption{Overview of the dataset presented in this paper.\label{tab:dataset}}
\tabletypesize{\scriptsize}
\tablehead{
\colhead{Instrument} & \colhead{Waveband} &  \colhead{Energy} & \colhead{Date}  & \colhead{No. of}\\
\colhead{}           & \colhead{}         &  \colhead{range}  & \colhead{range} & \colhead{observations\tablenotemark{a}}
}
\startdata
VERITAS              & VHE-gamma-ray            & $>$ 200\,GeV                       & 2009 - 2017 & 87\\ 
{\it{Fermi}}-LAT     & HE-gamma-ray             & 0.1 - 500\,GeV                 & 2008 - 2017 & 1045\tablenotemark{b}\\
{\it{Swift}}-XRT     & X-ray                    & 0.3 - 10\,keV                  & 2009 - 2017 & 25 \\
{\it{Swift}}-UVOT    & UV-optical               & 170 - 650\,nm\tablenotemark{c} & 2009 - 2017 & 232 \\
Tuorla               & Optical                  & $R$-band                             & 2003 - 2017 & 424\\
NOT                  & Optical\tablenotemark{d} & $R$-band                                 & 2014 - 2017 & 49\\
OVRO                 & Radio                    & 15\,GHz                            & 2008 - 2017 & 475\\
Mets\"{a}hovi            & Radio                    & 37\,GHz                            & 2002 - 2016 & 53\\
VLBA (MOJAVE)        & Radio                    & 15.3\,GHz                          & 2009 - 2016 & 10 \\
VLBA                 & Radio                    & 22.2 \& 43.1\,GHz                    & 2014        & 2 \\
\enddata
\tablenotetext{a}{We list here the number of flux points shown in Figure~\ref{fig:all-lightcurves} to give an indication of the sampling at each wavelength. For the VLBA observations, we just provide the number of images that were recorded.}
\tablenotetext{b}{Number of flux points in the 3-day binned light curve.}
\tablenotetext{c}{The UVOT data were taken with six different filters with central wavelengths of 544\,nm (V filter), 439\,nm (B filter), 345\,nm (U filter), 251\,nm (UVW1 filter), 217\,nm (UVM2 filter) and 188\,nm (UVW2 filter) \citep{2005SSRv..120...95R}.}
\tablenotetext{d}{The NOT provided polarization measurements at optical wavelengths.}
\end{deluxetable*}

In this work, we investigate the broadband emission of 1ES\,1215+303 using multiwavelength (MWL) observations (radio, infrared, optical, ultraviolet, X-ray and gamma-ray) covering the past decade, with a focus on the gamma-ray data. 
Given that one luminous gamma-ray flare has already been detected, we were interested in exploring the long-term temporal behavior of the source using observations from the {\sl Fermi}-LAT and VERITAS.

\section{Observations and Data Analysis} \label{sec:analysis}

An overview of the observations analyzed for this paper and of the instruments that made them is provided in Table~\ref{tab:dataset}.

\subsection{VHE Gamma-ray Data: VERITAS}
\label{sec:VERITAS}

VERITAS is sensitive to gamma rays in the energy range between $\approx\,$85\,GeV and $>$30\,TeV \citep{Park15}. 
It has a field-of-view (FoV) of $\approx\,$3.5$^\circ$. This makes it possible to observe simultaneously sources with small angular separation such as 1ES\,1215+303 and 1ES\,1218+304 (the angular distance between the two is $\approx\,$0.76$^\circ$). They have been monitored regularly since 2008 December. 
These observations were taken in ``wobble mode''~\citep{Fomin94} with the source (either 1ES\,1215+303 or 1ES\,1218+304) offset by 0.5$^{\circ}$ from the center of the FoV. 
The total exposure with 1ES\,1215+303 in the FoV between 2008 December and 2017 May (after quality selection, before dead-time correction, without accounting for the difference in sensitivity between observations on the two sources) amounts to 175.8\,h. 
The VERITAS results on this source between 2008 December and 2012 May were reported in \cite{Aliu13}, and those between 2013 January and 2014 May, including an extremely luminous flare, in \cite{Abeysekara17B21215}. 

The VERITAS data were analyzed using two independent packages \citep{Cogan08, Maier17}, and consistent results were obtained. 
Cuts on air shower image parameters optimized for each analysis package for a point source of 2\% to 10\% of the Crab Nebula flux with a power-law photon index between 2.5 and 3.0 \citep{Park15} were used. 

\begin{deluxetable}{cccc}[ht!]
\tablecolumns{4}
\tabletypesize{\scriptsize}
\setlength\tabcolsep{1pt}
\tablecaption{\footnotesize VERITAS observations of 1ES\,1215+303 from 2008 December to 2017 May. The VERITAS observing season runs from the end of the monsoon season ($\approx\,$September) until the start of the monsoon season the following year ($\approx\,$July) and is divided into periods called ``darkruns'' that are centered on the new moon.\label{tab:Vobs}}
\tablehead{
\colhead{Epoch} & \colhead{Exposure} & \colhead{Flux $>$200 GeV}& \colhead{Photon Index} \\ 
\colhead{} & \colhead{(hr)}  & \colhead{($\text{cm}^{-2} \; \text{s}^{-1}$)} & 
} 
\startdata
2008-2009	&	33.8	&	$<4.5\times 10^{-12}$ 		&	-	\\
2010-2011	&	41.9	&	$(8.0\pm0.9)\times 10^{-12}$ &	$3.6\pm0.4$	\\
2011-2012	&	17.5	&	$(2.8\pm1.1)\times 10^{-12}$ &	-	\\
\hline
2012-2013 non-flare	&	10.8	&	$(6.0\pm1.2)\times 10^{-12}$ &	$3.9\pm0.6$ \\
2013 Feb 07 (2)	&	0.5		&	$(5.1\pm1.0)\times 10^{-11}$ &	$3.7\pm0.7$ \\
\hline
2013-2014 non-flare	$^\dagger$ &	7.4	&	$<7.2\times 10^{-12}$ &	- \\ 
2014 Feb 08 (3)	&	0.9		&	$(5.0\pm0.1)\times 10^{-10}$ &	$3.1\pm0.1$ \\
\hline
2014-2015 non-flare	&	14.4	&	$(4.2\pm0.8)\times 10^{-12}$ &	$2.8\pm0.4$ \\
2015 Jan 17 (4)	&	0.9		&	$(5.3\pm0.5)\times 10^{-11}$ &	$3.0\pm0.2$ \\
\hline
2015-2016 non-flare	&	22	&	$(1.3\pm0.1)\times 10^{-11}$ &	$3.3\pm0.1$ \\
2016 Apr 09	(5)&	0.9		&	$(3.7\pm0.5)\times 10^{-11}$ &	$3.1\pm0.3$ \\
\hline
2016-2017 non-flare	&	24.6	&	$(8.0\pm0.8)\times 10^{-12}$ &	$3.9\pm0.3$ \\
2017 Mar 05 (6)	&	0.9		&	$(5.9\pm0.9)\times 10^{-11}$ &	$2.5\pm0.4$ \\
2017 Apr 01 (7) &	2.5		&	$(9.5\pm0.6)\times 10^{-11}$ &	$3.6\pm0.1$ \\
\enddata
\tablecomments{The enumeration in parenthesis after the date of a flare corresponds to the flare ID. We refer to Sections \ref{sec:increasing-flux} and \ref{sec:seds} for details on the flare ID, the simultaneity of observations with the {\sl Fermi}-LAT and HE enhanced activity. \\
$^\dagger$ We reanalyzed the 2013 -- 2014 season non-flare data and report the upper limit of those observations. }
\end{deluxetable}

We found that a power-law model $dN/dE = N_0 \left(E/E_0 \right)^{-\Gamma}$ provides a good fit to the VERITAS spectra
, where $dN/dE$ is the differential photon flux, $N_0$ is the flux normalization at energy $E_0$, $\Gamma$ is the photon index, and $E$ is the photon energy. 

The VHE gamma-ray fluxes and best-fit photon indices for 1ES\,1215+303 for different epochs are shown in Table~\ref{tab:Vobs}. 
In most cases, no significant difference was found between the photon index measured during flares and that averaged over the quiescent part of the corresponding season, with the exception of the hard spectrum VERITAS flare on 2017 March 05. 

\begin{figure*}[ht!]
\begin{center}
\includegraphics[width=\textwidth,angle=0]{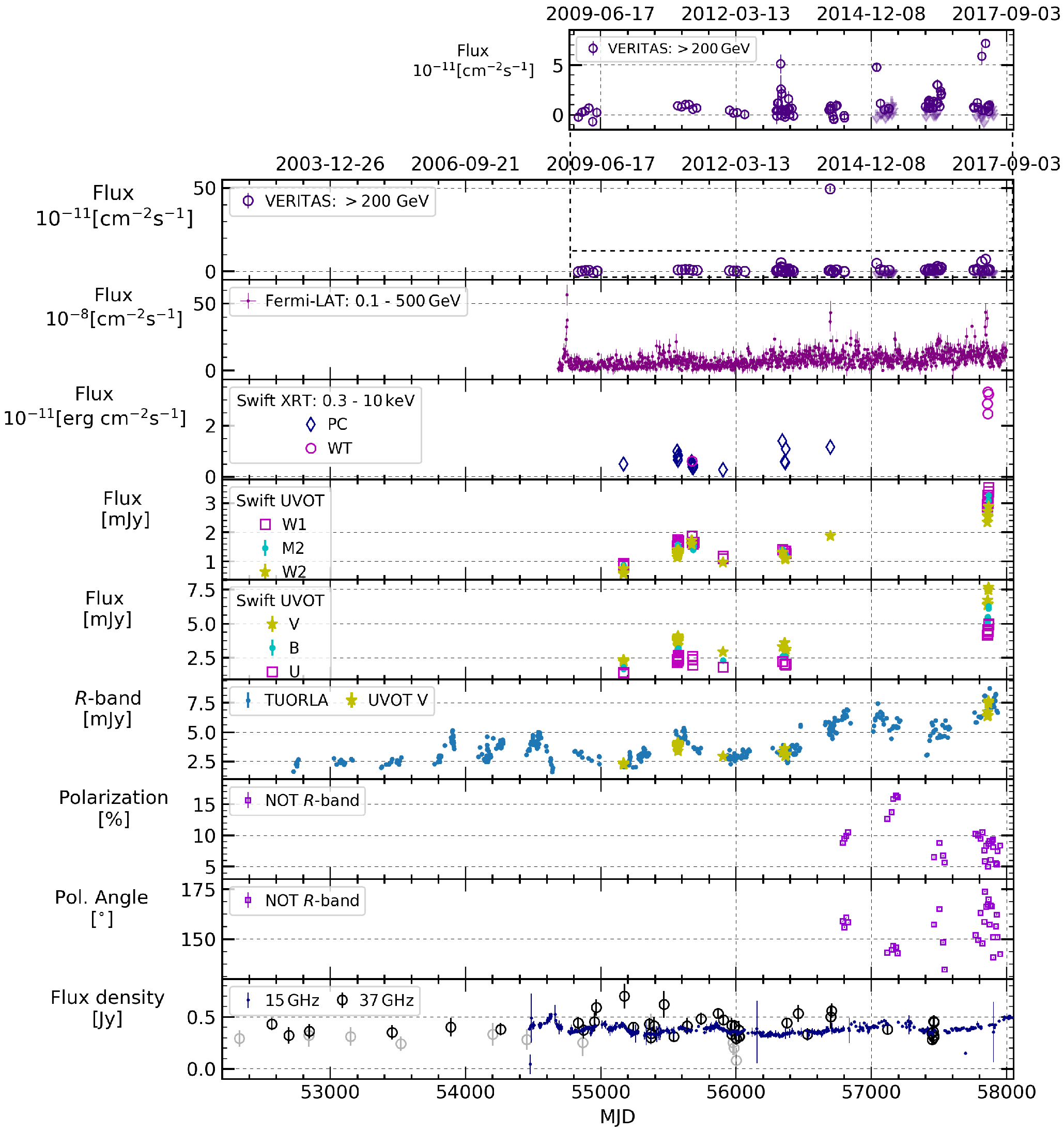}
\caption{The light curves for the various wave
  bands are shown in descending order of energy from the top to the bottom of the plot. A zoom is provided on the VERITAS data excluding Flare 3. For the XRT panel, the data taken in window-timing (WT) and photon-counting (PC) mode are plotted. For the radio panel, the 37~GHz data with signal to noise ratio S/N$<4$ are shown in gray. 
  \label{fig:all-lightcurves}}
\end{center}
\end{figure*}

\subsection{HE Gamma-ray Data: {\it{Fermi}} Gamma-ray Space Telescope $-$ LAT} \label{sec:fermi}

The Large Area Telescope, LAT, on board the {\it Fermi} Gamma-ray Space Telescope, covers the energy range from $\approx\,$20\,MeV to more than 500\,GeV \citep{Atwood09}. The main observation mode of the {\it Fermi}-LAT is survey mode during which the LAT scans the entire sky every 3 hours.
We analyzed the {\it Fermi}-LAT data from 2008 August 04 (MJD 54682.7), the start of the all-sky survey, up until 2017 September 04 (MJD 58001.0). 
The data were analyzed using the {\texttt{Fermi Science Tools}}\footnote{{\texttt{Version v10r0p5}; Instrument response functions {\texttt{P8R2\_SOURCE\_V6}}; the "source" class events were used.}}. 
We restricted the photon selection to those with energies between 100~MeV and 500~GeV that had zenith angle of less than 90$^{\circ}$ in order to reduce contributions from the Earth's limb. 
They consisted of photons in a circle of radius 10$^{\circ}$ centered on the position of 1ES\,1215+303, the region of interest (ROI). 
The data were modeled using the unbinned maximum likelihood fit method implemented in the {\it{Fermi}} Science Tools, \texttt{gtlike}. 
All of the sources from the third {\it Fermi}-LAT source catalog, 3FGL \citep{Acero15}, that lay within a radius of 20$^{\circ}$ of 1ES\,1215+303 were included in the background model to ensure that each source that could contribute photons to the ROI was modeled\footnote{The point spread function (PSF) of the {\it Fermi}-LAT is approximately 10$^{\circ}$ at 100\,MeV at the $95\%$ containment.}.

\begin{deluxetable}{ccccc}[ht!]
\tablecolumns{4}
\setlength\tabcolsep{1pt}
\tablecaption{\footnotesize {\it Fermi}-LAT flux and spectral shape of 1ES\,1215+303 from 2008 August 04, the start of {\it{Fermi}}-LAT science operations to 2017 September 04, the end of the period covered in this paper. The significance, flux and photon index are provided for the various different epochs listed in the table including each year (360-day bin), the flares (see Section \ref{sec:increasing-flux} to see how the flaring periods were defined) and for the yearly data excluding the flaring period(s). \label{tab:LATobs}}
\tabletypesize{\scriptsize}
\tablehead{
\colhead{Epoch} & \colhead{State} & Sig. &\colhead{Flux$_{>\text{0.1\,GeV}}$}& \colhead{$\Gamma$} \\ 
\colhead{} &  & \colhead{$\sigma$} & \colhead{$10^{-8}\text{cm}^{-2} \text{s}^{-1}$} & 
} 
\startdata
2008 Nov 17 - 2010 Aug 12 &     &  & & \\
\&                      & Low & 49.0 & $4.3\pm 0.3$ & $1.98\pm 0.03$ \\
2011 Apr 15 - 2012 Apr 09 &  &  &  & \\
2008 Aug 04 - 2009 Jul 30	& Total & 38.4 &	$ 5.3\pm 0.4 $	&	$1.94\pm 0.04$	\\
2008 Aug 04 - 2009 Jul 30	& Non-flare & 31.6 &$ 4.3\pm 0.4 $	&	$1.94\pm 0.04$	\\
\hline
2008 Oct 04 - 2008 Oct 17	& Flare\,1 & 26.4 &	$35.0\pm 3.5$ &	$1.92\pm 0.06$ \\
2009 Jul 30 - 2010 Jul 25	& Total & 29.3 &	$4.6\pm 0.5$ &	$2.01\pm 0.05$ \\
2010 Jul 25 - 2011 Jul 20	& Total & 40.9 &	$7.2\pm 0.5$ &	$1.97\pm 0.04$ \\
2011 Jul 20 - 2012 Jul 14	& Total & 32.8 &	$5.4\pm 0.5$ &	$2.00\pm 0.04$ \\
2012 Jul 14 - 2013 Jul 09	& Total & 47.0 &	$7.5\pm 0.5$ &	$1.92\pm 0.03$ \\
2013 Jul 09 - 2014 Jul 04	& Total & 54.0 &	$10.1\pm 0.6$ &  $1.94\pm 0.03$ \\
2013 Jul 09 - 2014 Jul 04	& Non-flare & 50.4 &	$10.0\pm 0.6$ &  $1.95\pm 0.03$ \\
2014 Jul 04 - 2015 Jun 29	& Total & 50.4 &	$8.7\pm 0.5$ &	$1.91\pm 0.03$ \\
2015 Jun 29 - 2016 Jun 23	& Total & 54.7 &	$9.1\pm 0.5$ &	$1.90\pm 0.03$ \\
2016 Jun 23 - 2017 Jun 18	& Total & 70.1 &	$12.0\pm 0.5$ &	$1.86\pm 0.02$ \\
2016 Jun 23 - 2017 Jun 18	& Non-flare & 63.7 &	$11.2\pm 0.5$ &	$1.88\pm 0.02$ \\
2017 Mar 25 - 2017 Apr 05	& Flare\,7 & 25.9 &	$25.2\pm 2.8$ &	$1.74\pm 0.06$ \\
2017 Apr 09 - 2017 Apr 16	& Flare\,8 & 18.9 &	$28.4\pm 4.0$ &	$1.83\pm 0.08$ \\
2017 Apr 15 - 2017 Apr 23 & Post-flare & 8.6 & $9.5\pm 2.3$ & $1.89\pm 0.34$ \\
\enddata
\tablecomments{Sig. stands for significance, while $\Gamma$ represents the power-law photon  index. 
We refer to Sections \ref{sec:increasing-flux} and \ref{sec:seds} for details on the flare ID, and the simultaneity of observations with VERITAS and GeV enhanced activity.}
\end{deluxetable}

Table \ref{tab:LATobs} shows the best-fit values for the  power-law spectral shape parameter and for the flux obtained for the different epochs, flaring, low state, post-flare, 360-day binned (approximately yearly), and 360-day binned outside flares (non-flare) results. The low state and post-flare states were defined using the Bayesian blocks method as described in Section \ref{sec:increasing-flux}. 

Systematic uncertainties were not included in the reported LAT data. They are estimated to be up to 10\%, based on the systematic uncertainties on the effective area and on the PSF\footnote{\url{https://fermi.gsfc.nasa.gov/ssc/data/analysis/LAT_caveats_p8r2.html}}.

\subsection{X-ray Data: Neil Gehrels {\it{Swift}} Observatory $-$ XRT} \label{sec:xrt}

The X-Ray Telescope \citep[XRT;][]{2000SPIE.4140...64B} on the \textit{Neil Gehrels Swift Observatory} 
is sensitive to photons with energies between 0.2 and 10\,keV \citep{Gehrels04, Burrows05}. There were 25 pointed \textit{Swift}-XRT observations within a $10^\prime$ radius of 1ES\,1215+303, 20 of which were taken in photon counting mode, and five in windowed timing mode. Only five observations were taken after 2013, one on 2014 February 9 (MJD\,56697) and four between 2017 April 15 (MJD\,57858) and 2017 April 23 (MJD\,57866), all of which were triggered by elevated VHE gamma-ray fluxes detected by VERITAS. 
The XRT data were initially processed using \texttt{xrtpipeline}\footnote{\texttt{HEASOFT v6.23}, \texttt{swxrtdas\_23Jan18\_v3.4.1} with calibrations from database \texttt{CALDB 20171113}.}. For subsequent spectral and temporal analysis, we used a circular source region of a radius of 20 pixels ($\approx\,47.2''$) and an annular background region with inner and outer radii of 70 and 120 pixels ($\approx\,2.75^{\prime}$--$4.72^{\prime}$), respectively, both centered on 1ES\,1215+303. We checked the count rate in the source region for each observation, and confirmed that the pile-up effect is negligible.

The X-ray spectrum was fit with an absorbed power-law model (\texttt{wabs*powerlaw}): 
\begin{equation}
\label{eqXspec}
\frac{dN}{dE} = e^{-N_H\sigma(E)} K \left( \frac{E}{1\;\text{keV}} \right)^{-\Gamma}, 
\end{equation}
where the column density of neutral hydrogen $N_H$ and the photoelectric cross-section $\sigma(E)$ describe the absorption component, and the normalization $K$ and photon index $\Gamma$ describe the power-law component. 
We fixed the column density of neutral hydrogen to $N_H = 1.74  \times10^{20} \;\text{cm}^{-2}$ taken from the Leiden/Argentine/Bonn (LAB) survey of Galactic HI \citep{Kalberla05}. 
The best-fit photon index, the energy flux between 0.3\,keV and 10\,keV, and the goodness of the fit for each observation is shown in Table~\ref{tab:Xspec} in Appendix~\ref{app:xrt}.

\subsection{Ultraviolet Data: Neil Gehrels {\it{Swift}} Observatory $-$ UVOT}
\label{sec:UVOT}

The Ultraviolet/Optical telescope \citep[UVOT;][]{2005SSRv..120...95R} on the \textit{Neil Gehrels Swift Observatory} made many observations of 1ES\,1215+303 during the time period under study in this paper.
Specifically, 232 images containing 1ES\,1215+303 in the field of view were available (31 with the V filter; 36 with the B filter; 40 with the U filter; 46 with the UVW1 filter; 42 with the UVM2 filter; 37 with the UVW2 filter) and they span the date range from 2009 December 03 (MJD 55168) to 2017 April 23 (MJD 57866). Since UVOT is co-aligned with the XRT, the temporal sampling of the observations from the two instruments is the same.
The counts from the source were extracted from a $5.0''$ (radius) aperture around the position of 1ES\,1215+303.
The background counts were estimated using the counts from four neighboring dark-sky regions, each having the same radius as the source region. The magnitude was then computed using the \texttt{uvotsource}\footnote{\texttt{HEASOFT v6.21}, \texttt{Swift$\_$Rel4.5(Bld34)$\_$27Jul2015} with calibrations from \citet{2011AIPC.1358..373B}.} tool.
The counts were first corrected for extinction following the procedure and using the $R_\textrm{v}[\equiv A(V)/E(B-V)]$ value of \citet{2009ApJ...690..163R}.
They were then converted to fluxes using the zero-point values for each of the UVOT filters from \citet{2008MNRAS.383..627P}. We used the values\footnote{The wavelength dependent coefficients $a$ and $b$ are defined according to $A_{\lambda}=E(B-V)[aR_\textrm{v}+b]$.} of $a$ and $b$ from \citet{2009ApJ...690..163R}, who computed them following the procedure of \citet{1989ApJ...345..245C}.
A value of 0.021 was used for $E(B-V)$ \citep{2011ApJ...737..103S}; this was accessed through the NASA/IPAC Extragalactic Database\footnote{\url{http://nedwww.ipac.caltech.edu}}.

\subsection{Optical Data} \label{sec:opt}

1ES~1215+303 was monitored in the $R$-band at the Tuorla Observatory over the past 15 years as part of the Tuorla blazar monitoring program \citep{Takalo08, Nilsson2018}. 
We show the long-term $R$-band flux density in Figure~\ref{fig:all-lightcurves}. \\

The source was monitored with the Nordic Optical Telescope (NOT). The ALFOSC instrument is used in the standard setup for linear polarization observations ($\lambda/2$ retarder followed by a calcite). The observations were performed in the $R$-band from 2014 to 2017 two to four times per month. The data were analyzed as in \cite{Hovatta2016} with a semi-automatic pipeline using standard aperture photometry and comparison stars procedures.

\subsection{Radio Data: VLBA} \label{sec:vlba}

1ES\,1215+303 was observed with the Long Baseline Observatory's Very Long Baseline Array (VLBA) at 22.2 and 43.1\,GHz on 2014 November 11\footnote{The results of these observations are publicly available at \url{http://whittierblazars.com/}} (observation code S7017E3). Approximately two hours of on-source integration time was recorded at each frequency, over a total time span of about seven hours. 
All observations used a 2\,Gbps recording rate in a dual-polarization configuration of eight 32\,MHz channels at matching frequencies in each polarization.

\begin{figure*}[ht!]
\begin{center}
\includegraphics[width=.45\textwidth,angle=0]{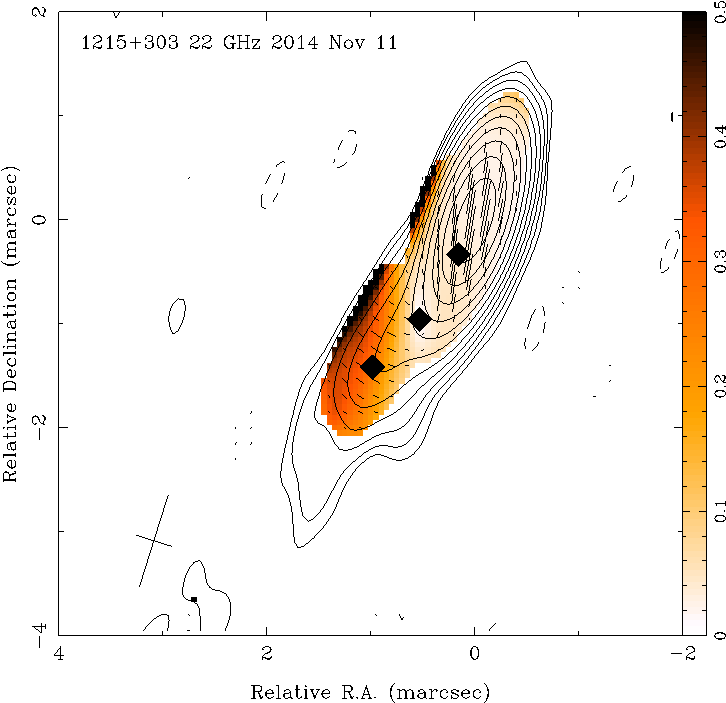}
   \put(-90,85){2}
   \put(-80,102){3}
   \put(-66,127){4}\hspace{20pt}
\includegraphics[width=.45\textwidth,angle=0]{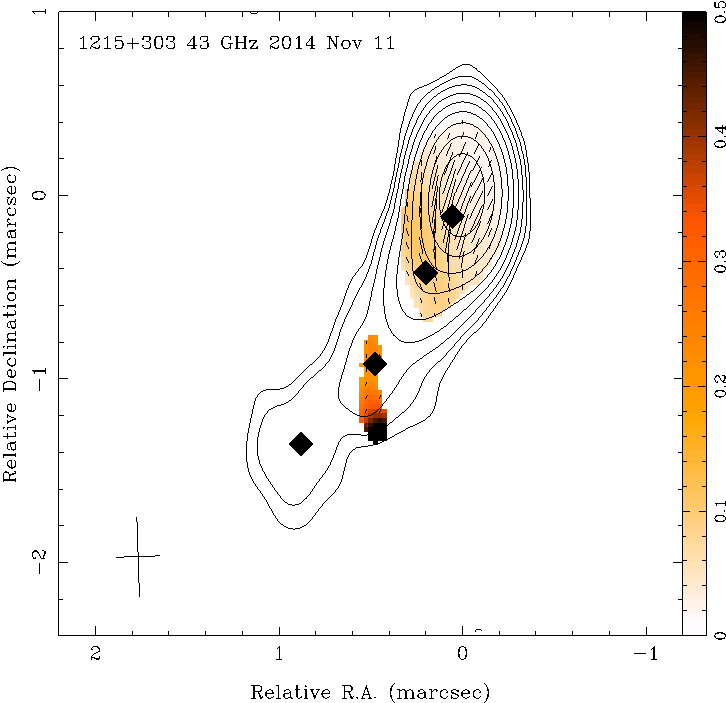}
   \put(-120,66){2}
   \put(-92,86){3}
   \put(-62,139){4}
\caption{{\it Left}: VLBA image at 22.2 GHz. Contours show total intensity, with the lowest contour at 0.129 mJy\,beam$^{-1}$, and subsequent contours factors of two higher. The peak flux density is 229 mJy\,beam$^{-1}$. The naturally-weighted beam size is 0.914 by 0.358\,mas at a position angle of the major axis of $-17.4^{\circ}$. Sticks show the magnitude of the linearly-polarized flux density (with a scale of 0.1 mas\,mJy$^{-1}$) and the direction of the EVPA.
The color scale indicates fractional polarization. {\it Right}: VLBA image at 43.1\,GHz. The lowest contour is 0.308 mJy\,beam$^{-1}$; the peak flux density is 152 mJy\,beam$^{-1}$. The naturally-weighted beam size
is 0.432 by 0.241\,mas at $1.9^{\circ}$. The polarized flux density scale of the sticks is 0.05 mas\,mJy$^{-1}$. The centers of the Gaussian jet components are shown as filled diamonds. The beams are shown in the bottom left-hand corner of each panel as a plus ``$+$''. \label{fig:vlba22_43}}
\end{center}
\end{figure*}
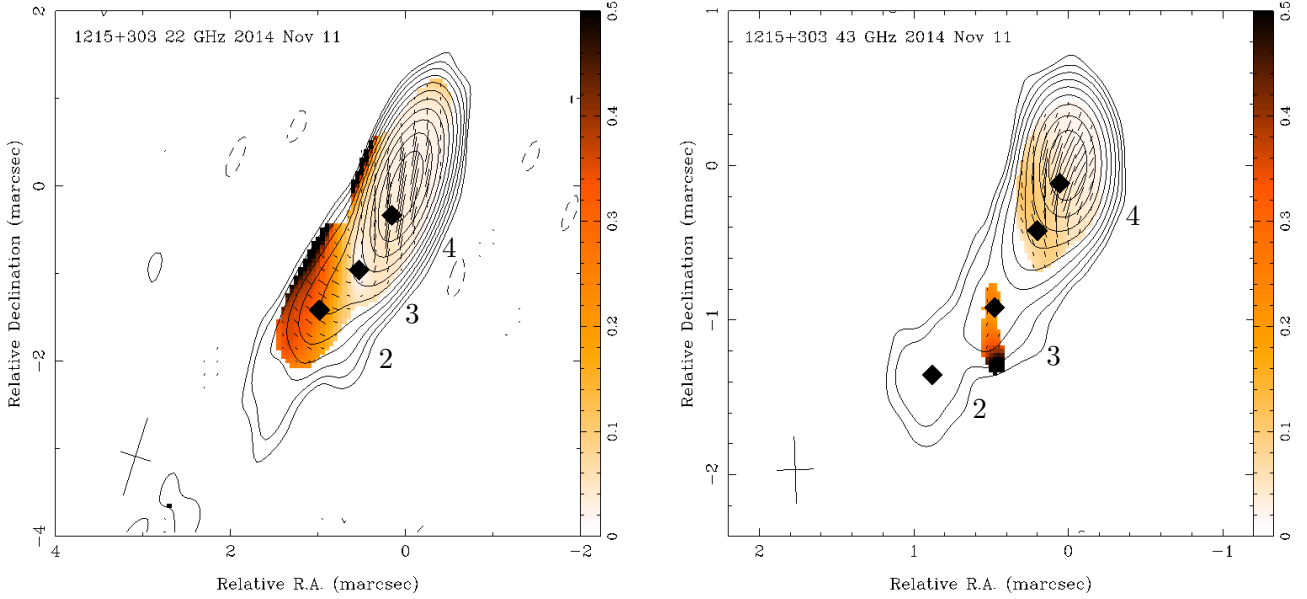

We used the \texttt{AIPS} software package \citep{Greisen03} for calibration and fringe-fitting of the correlated visibilities.
Calibration of the polarization response of the feeds (D-terms) was done through observations of standard calibrator sources.
Calibration  of  the  electric  vector position angle (EVPA) was done by comparison of  calibrator  sources  to  images  in  the  VLBA Boston monitoring  program, BU-BLAZAR\footnote{\url{https://www.bu.edu/blazars/VLBAproject.html}}, \citep{Jorstad16} or the Monitoring Of Jets in Active galactic nuclei with VLBA Experiments \citep[MOJAVE;][]{Lister19} databases\footnote{\url{http://physics.purdue.edu/astro/MOJAVE/sourcepages/1215+303.shtml}}.
Images were produced using \textit{clean} and \textit{self-calibration} in the \texttt{DIFMAP} software package \citep{Shepherd97}. All antennas were used for the 43.1~GHz image, and all except Saint Croix were used for the 22.2\,GHz image.

The 22.2\,GHz and 43.1\,GHz VLBA images are shown in Figure~\ref{fig:vlba22_43}. 
Both images exhibit fractional polarization increasing down the jet, relative to the core.
Circular Gaussian models were fit to the visibilities using the \texttt{modelfit} routine in \texttt{DIFMAP}.
In addition to the core, three jet components were detected at 22.2\,GHz, and four jet components were detected at 43.1\,GHz (with an additional component appearing between the innermost 22.2\,GHz component and the core). The centers of the Gaussian jet components are shown by filled diamonds on the VLBA images. The parameters of the Gaussian model components are tabulated in Table~\ref{tab:vlba}.

\begin{deluxetable}{cccccc}[ht!]
\tabletypesize{\scriptsize}
\tablecaption{VLBA 43.1, 22.2, and 15.3\,GHz Gaussian model
components. \label{tab:vlba}}
\tablehead{
 \colhead{Flux (Jy)} &  \colhead{$r$ (mas)} &  \colhead{P.A. ($^\circ$)} &  \colhead{$a^\dagger$ (mas)}  &  \colhead{Freq (GHz)} &  I.D. \\
 \colhead{(1)} &  \colhead{(2)} &  \colhead{(3)} &  \colhead{(4)} &  \colhead{(5)} &  \colhead{(6)}   
 }
\startdata
     0.127 &        0.03 &        $-16.1$ &       0.04 &   43.1 &   0 \\
     0.044 &         0.13 &          155 &       0.1 &   43.1 &   - \\
     0.014 &         0.47 &          155 &        0.2 &   43.1 &   4 \\
     0.003 &          1.04 &          153 &        0.30 &   43.1 &   3 \\
     0.003 &          1.62 &          147 &        0.39 &   43.1 &   2 \\
\midrule
     0.207 &        0.04 &        $-24.4$ &       0.02 &   22.2 &   0 \\
     0.038 &         0.37 &          155 &        0.11 &   22.2 &   4 \\
     0.008 &           1.1 &          151 &        0.30 &   22.2 &   3 \\
     0.004 &          1.72 &          145 &        0.25 &   22.2 &   2 \\
\midrule
0.265	& 0.03 &	323.1 &	0.03  & 15.3$^\ddagger$ & 0 \\
0.033	& 0.47 &	152.5 &	0.12 & 15.3 & 4 \\
0.011	& 1.06 &	150.3 &	0.2 & 15.3 & 3 \\
0.009	& 1.67 &	145.6 &	0.34 & 15.3 & 2 \\
0.013	& 16.20 &	143.5 &	4.41 & 15.3 & 1 
\enddata
\tablecomments{Columns: (1) flux density of the component, 
			(2) and (3) the distance ($r$) and the position angle (P.A.) of the center of the component relative to the origin of the image, 
            (4) the full width at half maximum (FWHM) of the circular Gaussian component, 
            (5) measurement frequency, (6) Identification number of features from (or consistent with) \citet{Lister19}. \\
$^\dagger$ The standard deviations of the best-fit Gaussian components are approximately 20\% of the FWHM beam dimensions.\\
$^\ddagger$ The 15.3~GHz data correspond to fits using all data from the 10 epochs observed between 2009 and 2016.}
\end{deluxetable}

1ES~1215+303 was also observed at 15.3 GHz with the MOJAVE program for 10 epochs between 2009 and 2016. 

\begin{figure*}[ht!]
\begin{center}
\includegraphics[width=.4\textwidth]{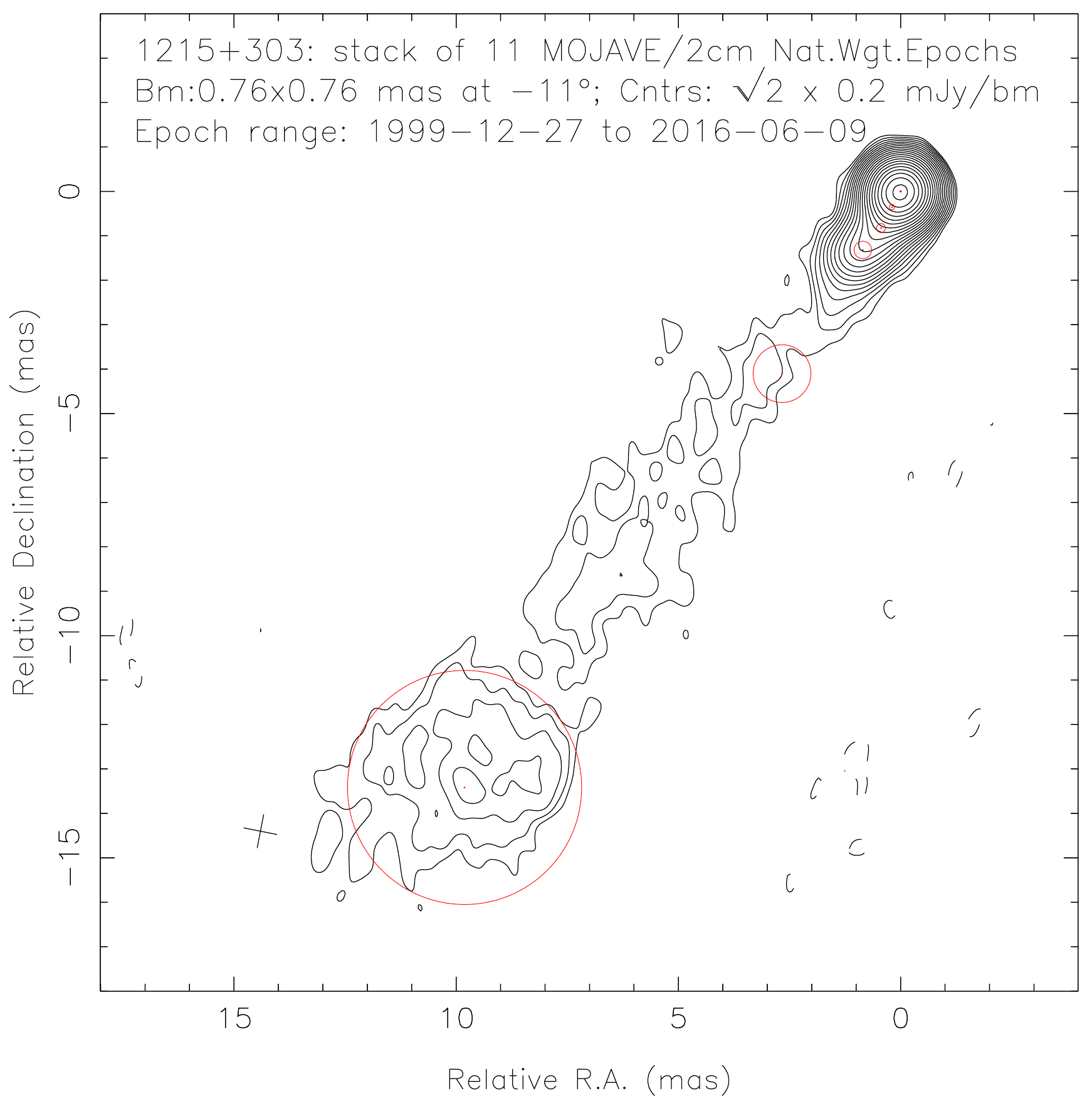}
   \put(-90,50){1}
   \put(-38,145){2}
   \put(-31,151){3}
   \put(-27,160){4}
\includegraphics[width=.555\textwidth]{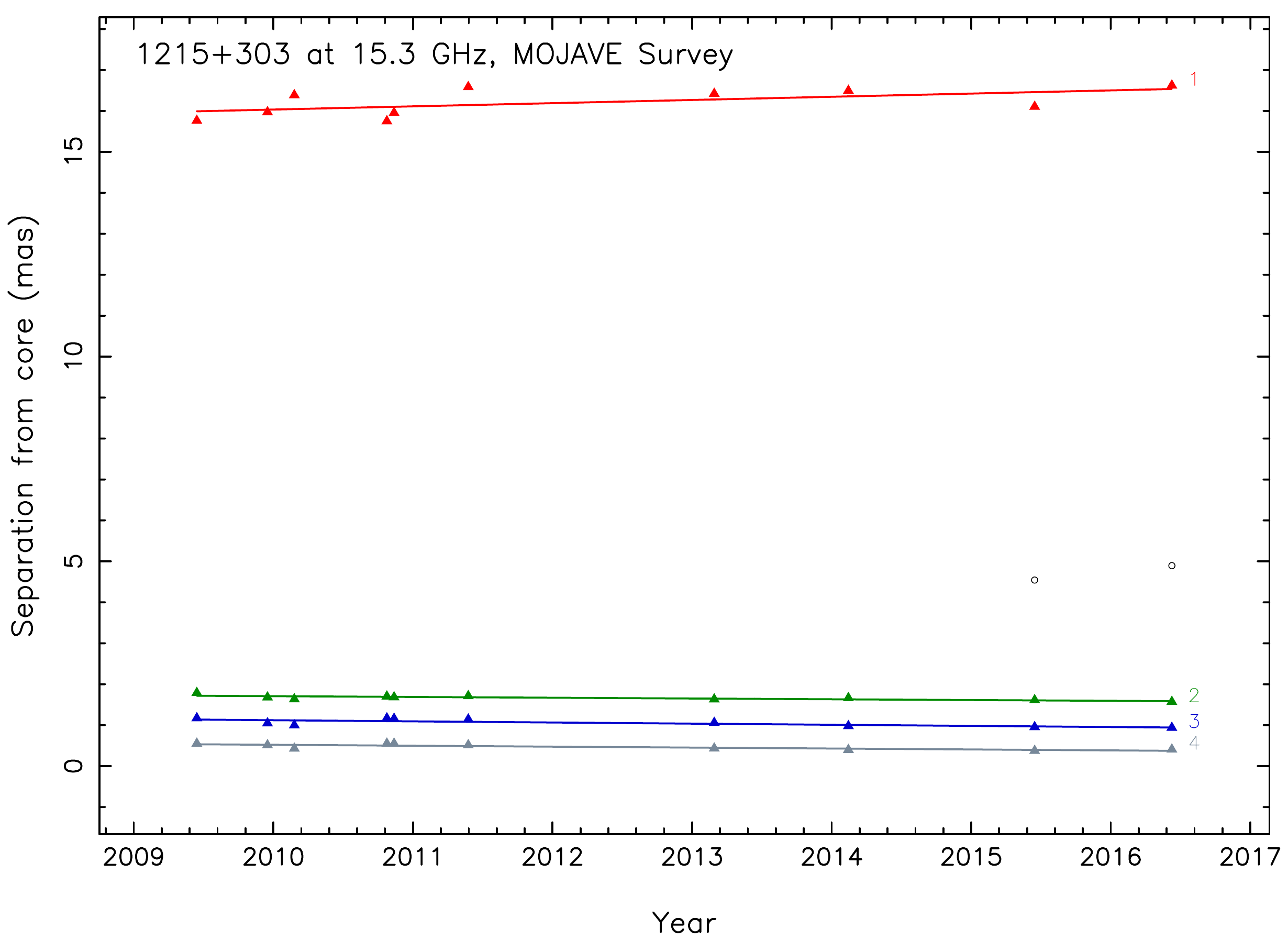}
\caption{{\it Left}: The stacked MOJAVE image with the five best-fit Gaussian components from the last epoch on 2016 June 9 overlaid. The standard deviations of the best-fit Gaussian components are approximately 20\% of the FWHM beam dimensions. The contours show the total intensity, starting at a baseline of 0.2 mJy beam$^{-1}$, and incrementing by factors of $\sqrt{2}$. Eleven images are stacked here including one from 27 December 1999 which is not shown on the plot on the right. The same circular restoring beam was used for all eleven images. It is shown at the half power level in the bottom left corner as a plus ``$+$''.
{\it Right}: The separation between components and the core at the time of each epoch of observation. The innermost three components (designated with number 2, 3, and 4) are quasi-stationary. Robust features which are cross-identified between more than 4 epochs are fitted assuming linear motion.
\label{fig:mojave}}
\end{center}
\end{figure*}

Emission features derived from a Gaussian model fit to the interferometric visibility data have been identified in the VLBA images at 15.3~GHz. The separations between these emission features and the core at the time of each epoch of observation are shown in the right panel of Figure~\ref{fig:mojave}, revealing three innermost emission features (components), referenced as 2, 3, and 4. Stationary features are typical in TeV HBLs, being present in the majority of these sources \citep{Kharb08, Hervet16, Piner18, Lico12}.
The mean and standard deviation of the angular separation between the three quasi-stationary components and the core over all epochs are $0.44\pm0.07$ mas, $1.04\pm0.09$ mas, and $1.64\pm0.06$ mas, as shown in Table~\ref{tab:vlba}. 
These three stationary components are also resolved in the 22.2 and 43.1-GHz images, and the positions of these three Gaussian components are consistent between the three frequencies. 
The fourth component observed at 15.3\,GHz is at a much larger distance from the origin of the images, in a position consistent with a very-long-baseline interferometry  (VLBI) stationary component found at 1.6 and 5\,GHz \citep{Giroletti06}.\\

The components 2, 3, and 4 show subluminal inward apparent speeds respectively of $0.170 \pm 0.036 \;c$, $0.246 \pm 0.055 \;c$, and $0.194 \pm 0.040 \;c$ estimated by MOJAVE. The fact that they have similar inward motions, all consistent with an inward speed of $0.2 \,c$, suggests that they are due to a downstream shift of the radio core. Indeed, if the three features are stationary shocks, a core shift predicts a similar inward motion for all of them.
Such a shift of the radio core can be explained by a slow increase of the jet power over years, which would increase the distance from the supermassive black hole (SMBH) where the jet becomes optically thin in radio. Such a slow power increase is supported by the multi-year increase of the gamma-ray and optical luminosities reported in Section \ref{sec:temporal}. Similar inward motions have been detected in other BL Lac sources by MOJAVE such as UGC 00773, 3C 66A, and Mrk 421 \citep{Lister19}.

Since the emission features are quasi-stationary, we show a stacked image of the 15.3 GHz intensity in the left panel of Figure~\ref{fig:mojave}. The five best-fit Gaussian components from the last epoch on 2016 Jun 9 are shown as red circles.

\subsection{Radio Data: Owens Valley Radio Observatory} \label{sec:ovro}

We show the radio flux density measured by the Owens Valley Radio Observatory (OVRO) at 15~GHz over the past decade (2008-2017) in Figure~\ref{fig:all-lightcurves}, where a total of 475 data points are presented. 
The procedure of the OVRO data reduction and calibration procedures can be found in \citet{Richards11}.

\subsection{Radio Data: Mets\"{a}hovi} \label{sec:metsahovi}

We also show the radio flux density measured by Mets\"ahovi Radio Observatory (MRO) at 37\,GHz in Figure \ref{fig:all-lightcurves}. The duration of the MRO data are longer than those from OVRO, but the sampling is generally more sparse. 
The MRO data reduction and analysis procedure can be found in \citet{Terasranta98}.
The radio data were also used in the SED modeling in Section \ref{sec:modeling}, providing constraints on the less variable jet component.

\section{Temporal studies} \label{sec:temporal}

\begin{figure*}[ht!]
\begin{center}
\includegraphics[width=\textwidth,angle=0]{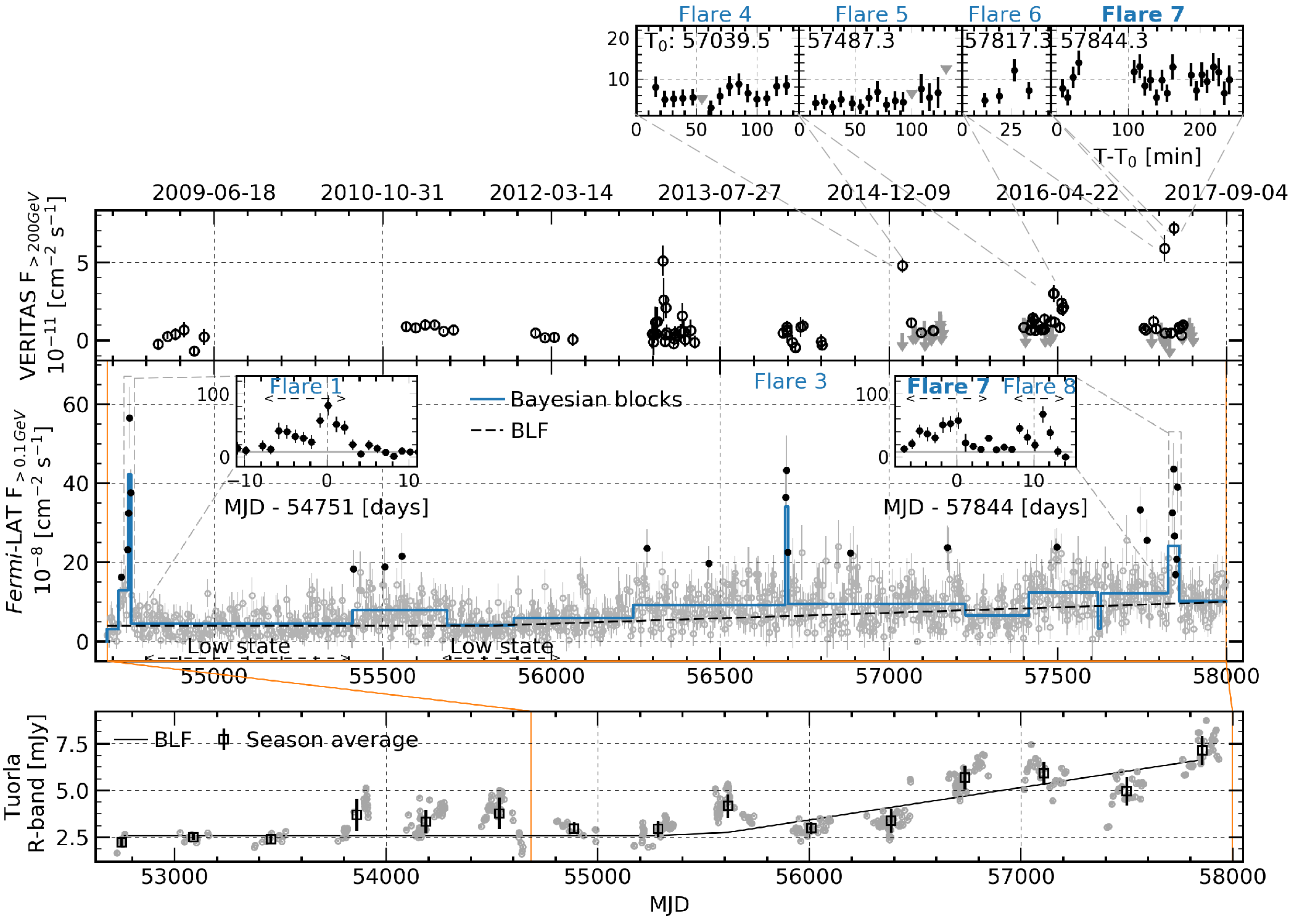}
\caption{The GeV-TeV full gamma-ray dataset. {\it Top}: The VERITAS light curve (above 200\,GeV and excluding Flare 3), with detailed zoom in the VHE flares down to the sub-hour timescales, from year 2015. \textsf{T}$_{\rm{0}}$ is in MJD. Upper limits in gray. {\it Middle}: {\it Fermi}-LAT 3-day light curve with daily zoom in the Flares 1, 7 and 8 (see text for details). Data points deviating $\geq 3\,\sigma$ from the broken linear function (BLF, dashed line) are shown in black. From these, only the points with two neighbors were used to define the four LAT flares. Bayesian blocks are shown in blue. These were used to define the low state of this source. {\it Bottom}: Tuorla light curve in gray with seasonal average in black. The last nine years are contemporaneous to the time range in the upper panels. \label{fig:gev-tev-lcs}}
\end{center}
\end{figure*}

In this section we describe various analyses that allow us to exploit the temporal richness of our dataset.

\subsection{The gamma-ray dataset}\label{sec:gamma}

We show the nightly VERITAS light curve integrated above 200\,GeV in the top panel of Figure~\ref{fig:all-lightcurves}. 
Flux values and their $1\,\sigma$ statistical uncertainties are shown only if the data result in a significance value of at least $2\,\sigma$, otherwise 95\% flux upper limits are shown. For the estimation of the integral flux points shown in the LAT light curve, each three-day dataset was subjected to the full likelihood analysis with the spectral parameters of all other sources in the ROI being frozen to those found in the global power-law likelihood analysis. For these short three-day exposures we found no preference for a curved spectral model so the 1ES\,1215+303 data were modeled as a power law.

As is discussed in detail in Sections \ref{sec:increasing-flux} and \ref{sec:seds}, 1ES 1215+303 flared a number of times at gamma-ray energies during the past decade, labeled Flares from 1 to 8. The names are assigned in chronological order to the gamma-ray flares, independently of whether they occur at HE or VHE. VERITAS gamma-ray flares were observed on six nights, Flares 2 to 7 (Table~\ref{tab:Vobs}). Two of these were found to have a counterpart at GeV energies, Flares 3 and 7. 
Flares 2 and 3 had a dedicated study reported in \cite{Abeysekara17B21215} while Flare 1 was analyzed along with 105 sources in \citet{Abdo10LC} and was not, therefore, subjected to a detailed, individual analysis. 
We focus on the unpublished observations and, in particular, on Flare 7 that occurred on 2017 April 01 since this is the only unpublished flare with simultaneous LAT-VERITAS data.

No strong intra-night variability on sub-hour timescales was observed in the light curves with 8-min binning intervals, as can be seen in the insets on the top panel of Figure~\ref{fig:gev-tev-lcs}.

\subsubsection{Increasing flux trend and definition of flares}
\label{sec:increasing-flux}

The second panel of Figure~\ref{fig:gev-tev-lcs} shows the {\it Fermi}-LAT 3-day binned light curve between 2008 August and 2017 September. Flux data and uncertainties are shown when a significance of at least $2\,\sigma$ was reached, otherwise 95\% upper limits are shown. In the following, low flux values were used instead of upper limits for the variability analyses.

\begin{deluxetable}{l|cccc}[ht!]
\tablecaption{ Results of the fit of the {\it Fermi}-LAT 3-day light curve and Tuorla averaged data.\label{tab:lat-tuorla}}
\tabletypesize{\tiny}
\tablehead{\noalign{\vskip4pt}
\colhead{Model} & \colhead{$a$} & \colhead{$b$} & \colhead{$t_{\rm break}$} & \colhead{$\chi^2$/d.o.f.}\\
 & \colhead{($\text{cm}^{-2} \, \text{s}^{-1}$\,MJD$^{-1}$)} & \colhead{($\text{cm}^{-2} \, \text{s}^{-1}$)} & \colhead{(MJD)} & \colhead{}
}
\startdata
\multicolumn{5}{c}{{\sl Fermi}-LAT}\\ \hline
Const.  &           NA                    & (5.57$\pm$0.14)$\,10^{-8}$ & NA        & 2361/1043 \\
Linear & (1.92$\pm$0.14)$\,10^{-11}$ & $-(1.02\pm0.08)\,10^{-6}$ & NA         & 1984.7/1042 \\
BLF & (2.75 $\pm$0.27)$\,10^{-11}$ & (4.00$\pm$0.20)$\,10^{-8}$ & 55834$\pm$134 & 1954.1/1041 \\ \hline
\multicolumn{5}{c}{Tuorla $R$-band}\\ \hline
Const.  &           NA                   & (2.92$\pm$0.25)$\,10^{-3}$ & NA & 102.8/13 \\
Linear & (5.46$\pm$1.10)$\,10^{-7}$ & $-(2.67 \pm 0.60)\,10^{-2}$ & NA & 35.4/12 \\
BLF & (1.73$\pm$0.44)$\,10^{-6}$ & (2.58$\pm$0.15)$\,10^{-3}$ & 55515$\pm$297 & 24.0/11 
\enddata
\tablecomments{For a linear function $ax+b$, $a$ is the slope and $b$ is the independent term. For a constant function $a$ is not applicable (NA). For the BLF, $a$ is the slope of the linearly increasing section, and $b$ is the value in the constant section.}
\end{deluxetable}

The LAT light curve comprises apparent flaring epochs on top of what looks like a variable baseline flux, which itself is not completely flat.
In order to characterize this baseline, we first fit the light curve to a constant flat line ($\chi^2_{\rm{red}}=2.26$)\footnote{The reduced $\chi^2$ is defined by $\chi^2_{\rm{red}}\equiv \chi^2$/d.o.f.}, to a linear function ($\chi^2_{\rm{red}}=1.90$) and to a broken linear function (BLF, $\chi^2_{\rm{red}}=1.88$).
A likelihood ratio test shows that the increasing linear function is preferred at the $19.4\,\sigma$ level to the constant fit and that the broken linear function (black dashed line in the second panel of Figure \ref{fig:gev-tev-lcs}) is preferred at the $5.5\,\sigma$ level over the increasing linear function. 
This broken line is composed of first a constant part given by $(4.0\pm0.2)\times 10^{-8}\;\text{cm}^{-2} \; \text{s}^{-1}$, consistent with the Bayesian blocks results (described below); and a linear function of slope $(2.8\pm0.3)\times 10^{-11}\;\text{cm}^{-2} \; \text{s}^{-1}$\,MJD$^{-1}$ which starts at the break point of MJD $55834\pm134$ (around September 2011). \\

A similar analysis was performed for the Tuorla $R$-band data averaged per season (black squares in the third panel of Figure \ref{fig:gev-tev-lcs}). It is found that a linear function is preferred at the $8.2\,\sigma$ level over a constant function, and that the broken linear function (in the same panel of Figure \ref{fig:gev-tev-lcs}) ($\chi^2_{\rm{red}}=2.2$) is preferred at the $3.4\,\sigma$ level over the linear function. 
The break point found for the Tuorla data is MJD 55515$\pm$297 (around November 2010), i.e., consistent with the LAT break time. See Table \ref{tab:lat-tuorla} for details on the results for both datasets. 
\cite{Lindfors2016} searched for long-term variability trends in Tuorla and 15\,GHz radio lightcurves from 2008 to 2013. No significant trend was found in radio or optical during this time period. This is not incompatible with our analysis, where the long term flux increase starts around the end of 2010 and become especially visible after 2013. The same study, however, reported to have found a decreasing or increasing trend for a number of other sources in the radio and optical bands.

In order to identify the LAT flaring epochs we performed a recursive fit on the data that deviated by no more than $3\,\sigma$ from the broken linear function (first method). 
This improved the fit ($\chi^2_{\rm{red}}=1.3$) and the results were consistent with those obtained before the $\geq 3\,\sigma$ points were excluded. 
The points that deviate by $\geq 3\,\sigma$ from this broken line are shown in black in Figure \ref{fig:gev-tev-lcs}. 
Out of these, only those with at least two neighboring flaring points, also above $3\,\sigma$, were used to define a LAT flare \citep{2015ApJ...814...35C}.
This method identified four {\sl Fermi} flares which we refer to as Flares 1, 3, 7 and 8.
The durations of the unpublished flares (1, 7 and 8) are provided in Table \ref{tab:LATobs} and they are plotted with one-day binning, in order to show their temporal structure in more detail, between the arrow edges in the insets of the second panel of Figure \ref{fig:gev-tev-lcs}. 
The peak day of Flare 7 is coincident between {\sl Fermi} and VERITAS observations, occurring on the night of 2017 April 01 (MJD 57844). This flare is annotated in bold font in the insets of Figure \ref{fig:gev-tev-lcs}). Details on the searches for simultaneous observations between the LAT and VERITAS are provided in Section \ref{sec:seds}.\\

The data were also divided into Bayesian blocks \citep[with a false positive rate, $p_0$, equivalent to $2.84\,\sigma$, see eq.\,(11) of][]{Scargle13}. 
The prior was chosen so that the flaring periods that we defined using the method described in Section~\ref{sec:increasing-flux} would be detected by this method, that is, Flares 1, 3, 7 and 8 in the case of the {\sl Fermi}-LAT.
The Bayesian blocks are in general agreement with the increasing trend, i.e., the flux of the blocks shows a mostly increasing trend starting approximately at the break time.
We used this method to find the periods during which 1ES\,1215+303 was in its ``low state'`(see Figure \ref{fig:gev-tev-lcs}) and also to define the 2017 post-flare state for the SED modeling (Section \ref{sec:modeling}).
The time periods of the different flux states can be found in Table \ref{tab:LATobs}.\\

The VERITAS light curve is characterized by a baseline at $\approx 2\%$ of the Crab Nebula flux. No preference was found for a long-term linear trend. The flares at this wavelength were selected when the photon flux rose above $10\%$ of the Crab Nebula flux. Between 2013 and 2017, these outbursts were observed at least once per year from 1ES\,1215+303.

\subsubsection{LAT spectrum and flux}
\label{sec:latSpectrumFlux}

\begin{figure}[ht!]
\begin{center}
\includegraphics[width=\linewidth,angle=0]{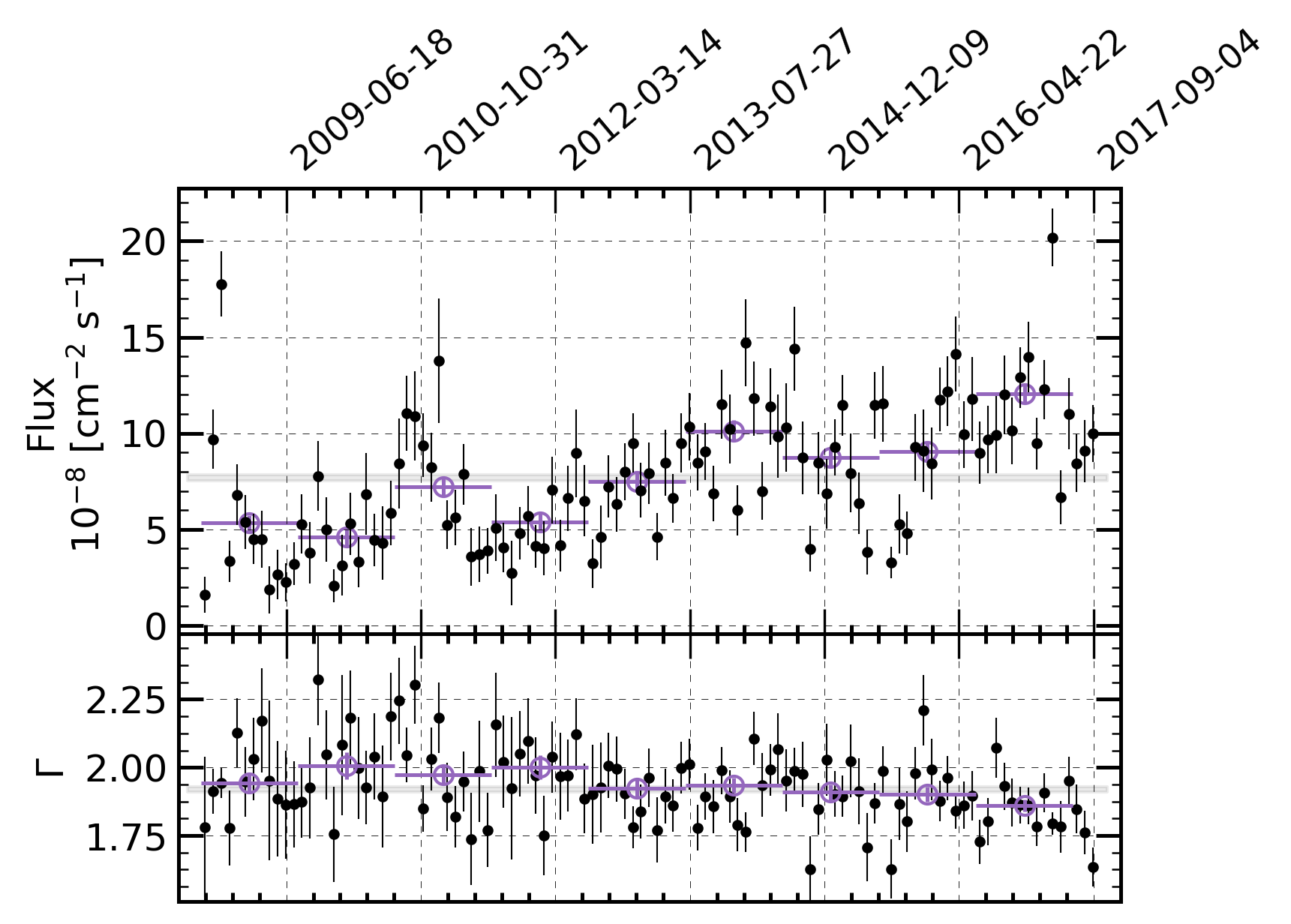}
\caption{{\it Top:} The 30-day binned (black points) and 360-day binned light curves (violet points). {\it Bottom:} Monthly spectral shape for the 30-day binned (black) and 360-day binned (violet) data. The gray shading in each of the two panels represent the value obtained for the entire 9-year data set.\label{fig:glc30+heph+hr60}}
\end{center}
\end{figure}

We analyzed each year of LAT data leaving both the flux and the photon index free so that the long-term evolution of these values could be investigated.
The analysis was also repeated with the flaring epochs excluded  (which are different from the yearly combined datasets only for those three years which included flaring episodes).
The results are shown in Figure \ref{fig:glc30+heph+hr60} and in Table \ref{tab:LATobs}.
The gray shading represents the values for the flux and the photon index obtained for the entire 9-year data set.

The nominal flux is sufficiently high to allow for binning on short timescales while still being able to extract significant information on the time evolution of the photon indices. Black points in the top and middle panels of Figure \ref{fig:glc30+heph+hr60} represent the monthly fluxes and photon indices respectively.
The data was also analyzed in 60-day bins to calculate the hardness ratio (HR) between two energy ranges, 0.1--1\,GeV and 1--500\,GeV. This analysis did not show significant changes in the HR for this source.

\begin{figure}[ht!]
\begin{center}
\includegraphics[width=.8\linewidth,angle=0]{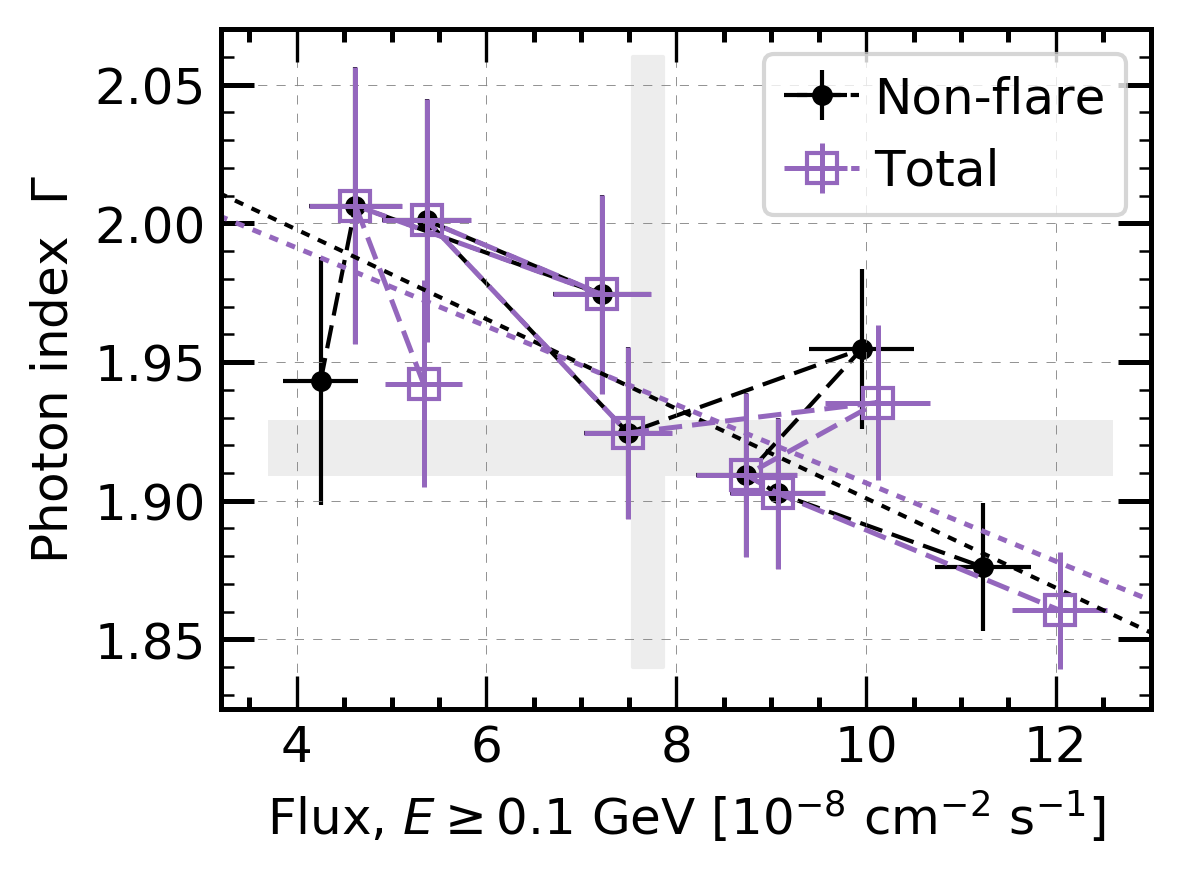}
\caption{Power-law photon index, $\Gamma$, against flux for the 360-day binned {\it Fermi}-LAT data. The violet square points show the average value per bin, while black points show the non-flaring state values. Dotted lines show the results of the linear fits. The 360-day light curve and photon indices against time are shown in Figure \ref{fig:glc30+heph+hr60} for the total data, in violet as well. The dashed lines join the data chronologically, going approximately from left to right, from where the long-term brightening and hardening can be visualised. \label{fig:index-flux}}
\end{center}
\end{figure}

There is strong evidence for a long-term hardening of this source, reaching the $5.0\,\sigma$ level with the 30-day binned data, ($4.7\,\sigma$ including trials factor by having looked at the 30, 60 and 360-day binned data); $3.6\,\sigma$ level for the yearly data bins and $3.2\,\sigma$ outside flares with this same binning. We observe a long-term brightening at this binning as well, reaching the $12.8\,\sigma$ level for the yearly data bins, and $13.4\,\sigma$ outside flares.
No photon index\,-\,flux or HR\,-\,flux correlation was observed for the 30-day or 60-day binned data, respectively. 
For the 360-day binned data, however, a Pearson correlation parameter of $-0.86$ between the photon index and the flux is obtained for the total data set (violet points in Figure \ref{fig:index-flux}), and a value of $-0.74$ for the non-flare data (black points in the same figure). 
A likelihood ratio test shows a $3.4\,\sigma$ preference, including trials factor (by having looked at the 30, 60 and 360-day binned data), for a linearly decreasing dependence over a constant between the photon index and the flux; which indicates a possible overall ``harder-when-brighter'' trend in this source. 
The yearly data outside flares also showed a preference at the $2.8\,\sigma$ level for a linearly decreasing dependence over a constant.  
These data, as well as the linear fits, are shown in Figure \ref{fig:index-flux} and the details of the fit parameters can be found in Table \ref{tab:index-flux} in Appendix \ref{app:yearlyfits}.
This ``harder-when-brighter'' trend has been observed in the {\it Fermi}-LAT data for flat-spectrum radio quasars and low-frequency-peaked BL Lacs \citep{2010ApJ...710.1271A}. We did not find any photons with $E > 50$\,GeV associated with any of the flares. The highest energy LAT photon detected had an energy of 466\,GeV and was detected on 2011 May 01 during a relatively high state of the source that lasted approximately 13 months.

\subsection{Multifrequency flux-flux cross-comparison and cross-correlations}
\label{sec:fluxflux}

\begin{figure*}[ht!]
\begin{center}
\includegraphics[width=.9\textwidth,angle=0]{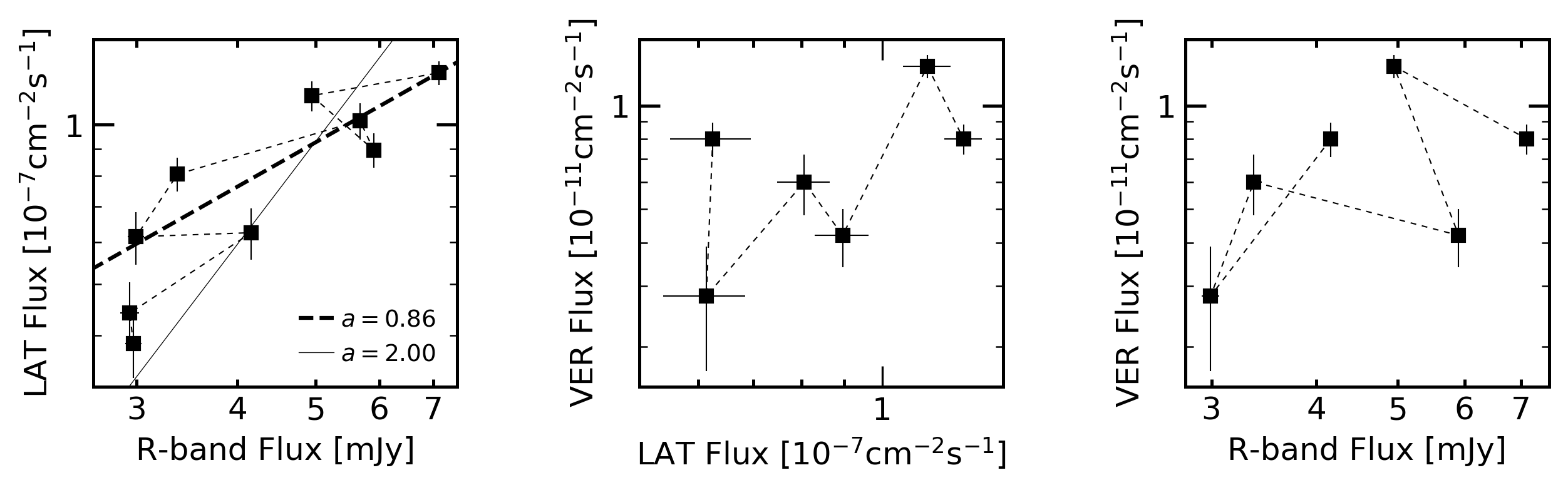}
\put(-432,37){{\tiny S1}}\put(-433,55){{\tiny S2}}\put(-387,68){{\tiny S3}}
\put(-432,77){{\tiny S4}}\put(-420,95){{\tiny S5}}\put(-352,106){{\tiny S6}}
\put(-348,94){{\tiny S7}}\put(-382,119){{\tiny S8}}\put(-336,113){{\tiny S9}}
\put(-263,106){{\tiny S3}}\put(-264,50){{\tiny S4}}\put(-235,94){{\tiny S5}}
\put(-211,68){{\tiny S7}}\put(-203,119){{\tiny S8}}\put(-188,97){{\tiny S9}}
\put(-80,106){{\tiny S3}}\put(-103,49){{\tiny S4}}\put(-102,94){{\tiny S5}}
\put(-30,68){{\tiny S7}}\put(-47,125){{\tiny S8}}\put(-22,96){{\tiny S9}}

\caption{Seasonal flux-flux diagrams for VERITAS, the {\it Fermi}-LAT and Tuorla ($R$-band) energy ranges (in logarithmic scale). The data is labeled from Season~1 (S1), in 2009, to Season~9 (S9), in 2017. The dotted lines join the data chronologically, going approximately from left to right due to the long-term brightening observed in the GeV and optical light curves. The dashed line represents the fit to the expression $\log_{{10}} (F_{\rm{LAT}})=a\,\log_{{10}} (F_{\rm{opt}})-b$. The solid line is the fit to the same expression with $a=2$. \label{fig:ff-year}}
\end{center}
\end{figure*}
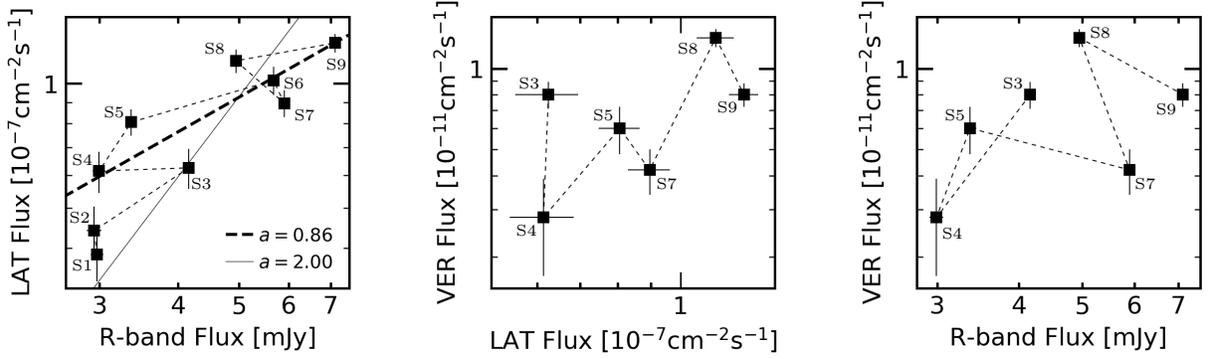

Attempts to search for flux-flux correlations using short time bins failed due to large uncertainties. Furthermore, the cross-correlation function analyses performed showed no evidence for significant inter-band correlation for the data shown in Figure \ref{fig:all-lightcurves} (see Section \ref{sec:zdcf} for details). 
We therefore performed a likelihood analysis of the LAT data using the $R$-band seasonal intervals (when the source was visible to optical telescopes), and analyzed the VHE gamma-ray data from the quiescent state for each year, shown in Table \ref{tab:Vobs}. 
The VERITAS data were taken between 2010 and 2017 and thus comprise seven data points, whereas the LAT data start in 2008, and therefore comprise 9 years of data, that is nine data points. 
The seasonal flux-flux correlations which result from these analyses are shown in Figure~\ref{fig:ff-year}, in logarithmic scale. 
The least-squares fits and Pearson correlation coefficients can be found in Table~\ref{tab:yearlycorr} for the logarithms of the seasonal fluxes for each set of energy bands. 
A strong long-term correlation between the optical and HE gamma-ray bands is found. 

We fitted the (GeV, optical) points with the expression
$$\log_{{10}} (F_{\rm{LAT}})=a\,\log_{{10}} (F_{\rm{opt}})-b$$ 
(dashed line in Figure~\ref{fig:ff-year}), yielding a slope $a=0.86 \pm 0.21$ and $b=5.05\pm 0.49$ with a $\chi^2/\rm{d.o.f.}=41/6$, and Pearson correlation coefficient of $0.86$. The uncertainties on $a$ and $b$ are obtained after having re-scaled the measurement uncertainties to $\chi^2/\rm{d.o.f.}=1$.\\

\begin{deluxetable}{l|cccc}[ht!]
\tablecaption{Seasonal flux logarithm correlations. \label{tab:yearlycorr}}
\tabletypesize{\scriptsize}
\tablehead{
\colhead{Energy bands} & \colhead{Pearson corr.} & \colhead{Linear fit\tablenotemark{*}} & $\chi^2/d.o.f.$ \\ 
\colhead{}  & \colhead{coefficient} & \colhead{slope} &  }
\startdata
  LAT - Optical & 0.86 & 0.86$\pm$0.21 & 41/6 \\ 
  VERITAS - LAT & 0.59 & 0.63$\pm$0.62 & 43/3 \\ 
  VERITAS - Optical & 0.44 & 0.06$\pm$0.80 & 54/3 \\ 
\enddata
\tablenotetext{*}{Uncertainties scaled to $\chi^2/d.o.f.$.}
\tablecomments{The linear fit slope corresponds to $a$ in a fit to: $\log(f_1)=a \,\log(f_2)+b$, where $f_1$ and $f_2$ are the seasonal fluxes in two different energy bands.}
\end{deluxetable}
 
 To our knowledge, this is the first time that such a strong global GeV-optical correlation has been observed over such an extended period of time (more than nine years).
The optical emission most likely comes from the synchrotron process and if the gamma-ray photons originate from inverse Compton scattering (ICS), this strong, almost linear ($a=0.86$) correlation is consistent with a long-term variability induced by changes of the Doppler factor or magnetic field of the emitting zone, considering a synchrotron-self-Compton (SSC) scenario. It is also consistent with gamma-ray emission originating from inverse-Compton scattering on an external photon field \citep[e.g.][]{2011MNRAS.410..368B}. \\

In a SSC scenario, if a change in the number of emitting particles is the cause of the long-term variability, this would induce a quadratic flux-flux correlation ($a = 2$ line in Figure \ref{fig:ff-year}) between the optical and the gamma-ray data. However, a slope of $a = 2$ is found to be disfavored at the $5.4\,\sigma$ level.  
If, instead of $\chi^2/$d.o.f. re-scaling, we add quadratically a source variability (of $\approx 30\%$), obtained from the excess variance analysis per season as in Section \ref{sec:fluxdist}, we obtain $a=0.83 \pm 0.33$, which would be preferred over $a=2$ at the $3.6\,\sigma$ level.\\ 

No evidence for a clear correlation was found between the HE and VHE bands. A weaker correlation is found between the VHE and the optical bands. 
No long-term correlation was observed between the OVRO data (15\,GHz) and the optical data or the gamma-ray data.

\subsection{Flux distributions and variability} \label{sec:fluxdist}

\begin{figure*}[ht!]
\begin{center}
\includegraphics[width=.367\linewidth,angle=0]{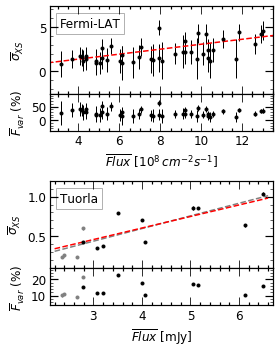}\hspace{1.cm}
\includegraphics[width=.5\linewidth,angle=0]{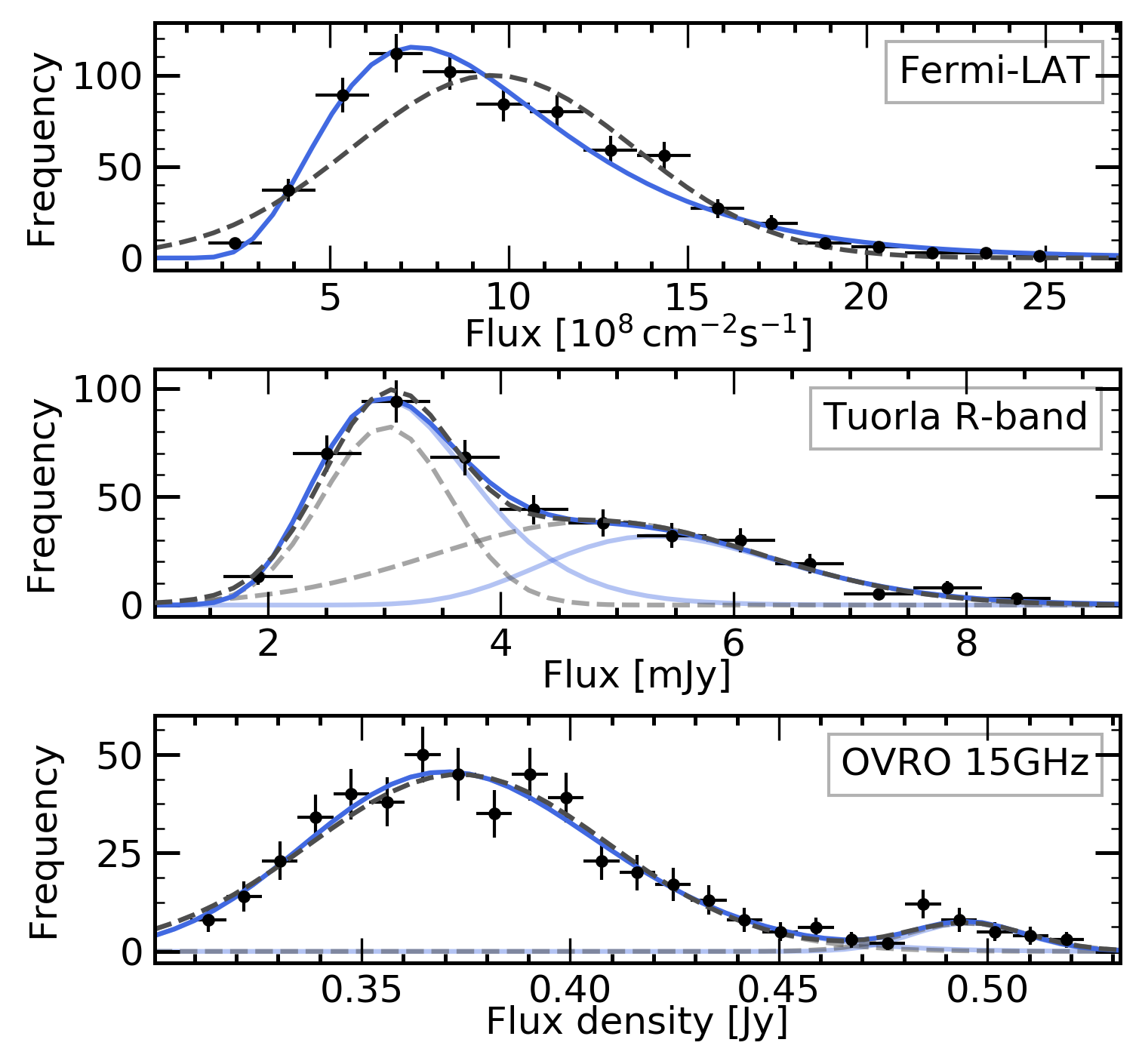}
\caption{{\sl Left:} The excess variance ($\overline{\sigma}_{XS}$) and variability amplitude ($\overline{F}_{\rm var}$) for the {\sl Fermi}-LAT and Tuorla data. {\sl Right:} LAT, Tuorla and OVRO flux distributions. The (bi)log-normal best fit is shown in solid light blue lines and the (bi)normal in dashed gray lines. The components of the bi-functions are shown in lighter blue for the bi-log-normal and in lighter gray for the normal function.\label{fig:fdist}}
\end{center}
\end{figure*}

In this section we analyze the flux distributions of the {best sampled} light curves from our observing campaign, namely, the OVRO, Tuorla, LAT 3-day binned and VERITAS data. 
These light curves are probed in order to search for log-normality in the distributions of their fluxes.

\begin{deluxetable*}{l|cc|ccc|cc|ccc}[ht!]
\tablecaption{Widths ($\sigma$) and goodness of fits ($\chi^2_{\rm{red}}$) for normal, bi-normal, log-normal and bi-log-normal fits to the LAT, Tuorla and OVRO flux data. \label{tab:bi-log-normal}}
\tabletypesize{\scriptsize}
\tablehead{
\colhead{Dataset} & \multicolumn{2}{c}{normal} & \multicolumn{3}{c}{bi-normal} & \multicolumn{2}{c}{log-normal} & \multicolumn{3}{c}{bi-log-normal}\\
 & \colhead{$\sigma$} & \colhead{$\chi^2_{\rm{red}}$} & \colhead{$\sigma_1$} & \colhead{$\sigma_2$} & \colhead{$\chi^2_{\rm{red}}$} & \colhead{$\sigma$} & \colhead{$\chi^2_{\rm{red}}$} & \colhead{$\sigma_1$} & \colhead{$\sigma_2$} & \colhead{$\chi^2_{\rm{red}}$} }
\startdata
  VERITAS & 0.38$\pm$0.05 & 0.76   & -                             & -                             &   -   & 0.63$\pm$0.06 & 0.97 & -             & -             & - \\
  LAT     & 3.9$\pm$0.3   & 4.12   & -                             & -                             &  -    & 0.43$\pm$0.02 & 1.42 & -             & -             & - \\
  Tuorla  & -             & -      & 0.5$\pm$0.1                   & 1.4$\pm$0.3                   & 1.48  & -             & -    & 0.22$\pm$0.02 & 0.19$\pm$0.04 & 1.08 \\
  OVRO    & -             & -      & (3.6$\pm$0.2)$\times 10^{-2}$ & (1.5$\pm$0.4)$\times 10^{-2}$ & 0.67 & -             & -     & (9.5$\pm$0.4)$\times 10^{-2}$ & (2.7$\pm$0.6)$\times 10^{-2}$ & 0.82 \\
\enddata
\tablecomments{The log-normal function is given by $f(x)=\frac{N}{x\sigma \sqrt{2\pi}}\exp\left[-\frac{(\log{x}-\mu)^2}{2\sigma^2}\right]$. A dash in a given column indicates that the particular function was not fit to that dataset.}
\end{deluxetable*}

This behavior has been studied in other blazars, such as BL\,Lacertae \citep{Giebels09}, 1ES\,1011+496 \citep{Sinha17} and a population of bright {\it Fermi} blazars \citep{Shah18} as well as in other accretion-powered systems \citep{3LACAckermann15}. 
Log-normal distributions have the property that their means and fluctuations behave linearly on average, and are of interest since they have multiplicative rather than additive properties \citep{Aitchison1973}. 

In order to estimate the fluctuations in the source flux that are not due to Poisson noise, the excess variance, $\sigma_{XS}$, was calculated. 
We binned the flux data points shown in Figure \ref{fig:all-lightcurves} in segments of equal duration and ensured that each bin contained at least 20 measurements of flux, excluding the flares. 
The excess variance $\sigma_{XS}^2 = \frac{1}{N}\Sigma^{i=1}_{N}(x-\overline{x})^2-\overline{\sigma^2_i}$ \citep[][Section B]{Vaughan03}  and the variability amplitude $\overline{F}_{\rm var}$ \citep{Vaughan03} are shown as a function of the flux arithmetic mean in the left hand side of Figure \ref{fig:fdist}. 
For the LAT data we obtain $\sigma_{XS}\propto (0.25\pm 0.05)\,\overline{\rm{Flux}}$ ($\chi^2_{\rm{red}}=$0.66) and a Pearson correlation coefficient, $\rho$, of 0.54. 
The Tuorla data are more sparsely sampled than the LAT data so some of the bins contain fewer than 20 flux measurements (gray points in the bottom left-hand panels of Figure \ref{fig:fdist}). 
A linear fit to these data outside of the low states yields $\sigma_{XS}\propto (0.15\pm 0.05)\,\overline{\rm{Flux}}$ ($\chi^2_{\rm red}=$172.5, $\rho=0.74$), while a linear fit to the total data set results in $\sigma_{XS}\propto (0.16\pm 0.04)\,\overline{\rm{Flux}}$ ($\chi^2_{\rm red}=$134.4, $\rho=0.80$). 
A similar analysis on the OVRO data did not show significant correlation ($\rho=-0.20$). 
It was not possible to perform this analysis on the VERITAS data due to their sparsity.

The flux distributions of the {\it Fermi}-LAT, Tuorla and OVRO 15\,GHz data, and their best fits to the (bi)log-normal (solid light blue) and (bi)normal (gray dashed) functions are shown in the right-hand panels of  Figure~\ref{fig:fdist}. Both bi-functions consist of two components each, which are shown in lighter colors in the same figure.
In the case of the LAT data, flaring states, as they were defined in the previous section, were excluded so as not to favor the log-normal fit (due to a possible bias produced by the elongated tail). 
A Shapiro-Wilk test on the LAT data rejects the normal distribution with a p-value of $4.2\times 10^{-16}$ and a test statistic of $w=0.87$ \citep{Shapiro1965}.  
The $\chi^2$ of the fits improve after Poisson noise reduction was applied during faint epochs, reaching the best fit when only data with significance above $3\,\sigma$ were included (approximately 60\% of the data below $3\,\sigma$ are located within the low state defined in Section~\ref{sec:increasing-flux}). 
The distribution of these data is shown in the top right-hand panel of Figure~\ref{fig:fdist}. The results of fits to normal ($\chi^2/$d.o.f.$=49.4/12$) and log-normal ($\chi^2/$d.o.f.$=17.0/12$) functions shown in the same figure, are presented in Table~\ref{tab:bi-log-normal}, where it is observed that the log-normal function provides a much better fit. 
The middle and bottom panels of the same figure show the Tuorla and OVRO flux distributions, respectively, where, contrary to the LAT data, no periods were excluded on the basis of the flux state of 1ES\,1215+303. 
This is because of the relative sparsity in the sampling of these light curves. 
We observe a double-peaked structure in their flux distributions, possibly due to the fact that both quiescent and flare data are included, or due to the presence of a brighter second quiescent state.
The bi-log-normal function does not provide a clear improvement to the fit with respect to the bi-normal function in the case of the Tuorla and OVRO data (see Table~\ref{tab:bi-log-normal}).
The two states of the Tuorla distributions are consistent with the states before and after the break time calculated in Section~\ref{sec:increasing-flux}. The bi-normal fit results of the OVRO distributions are consistent with the flux density of the states interpreted as quiescent and flaring components by \cite{Liodakis2017}, which includes data up to February 2016 for this source.
Two log-normal states were also previously observed at the IR-optical wavelengths in FSRQ PKS\,1510-089 \citep{Kushwaha16}. 
An analogous analysis performed on the VERITAS data outside flares did not show evidence for a preference for a normal over a log-normal function (see Table~\ref{tab:bi-log-normal} for the $\chi^2_{\rm{red}}$ values).

\subsection{ZDCF} \label{sec:zdcf}

\begin{figure}[ht!]
\begin{center}
\includegraphics[width=0.5\textwidth,angle=0]{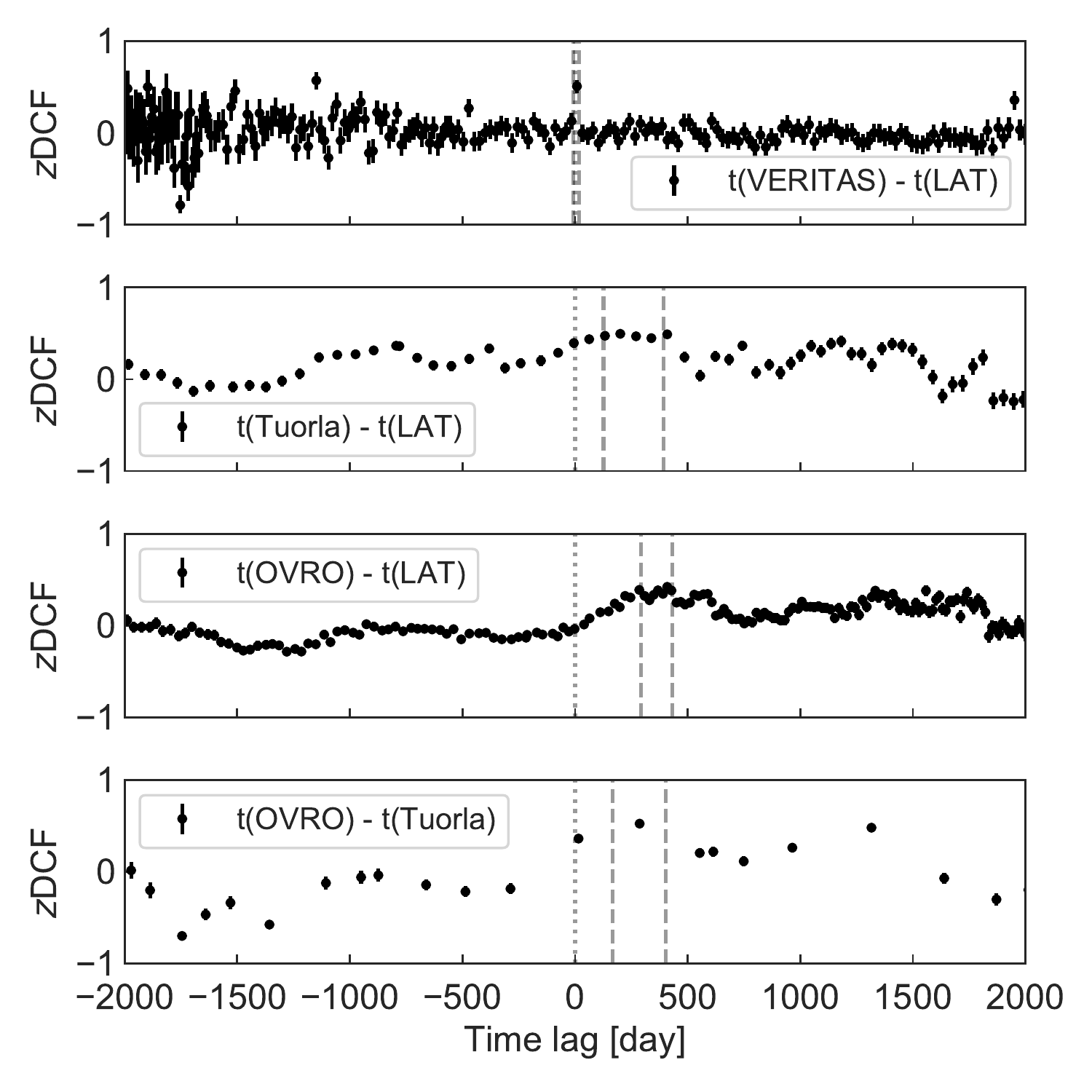}
\caption{The $Z$DCFs between light curves measured at different wavelengths. The pair of wavelengths in each panel is shown in the legend. A positive time lag ($t(X) - t(Y) >0$) between band $X$ and band $Y$ means the emission in band $X$ lags behind that in band $Y$. The vertical dotted lines show the time lag of zero, and the vertical dashed lines show the $1\,\sigma$ confidence interval around the maximum-likelihood peak time lag. \label{fig:ZDCFs}}
\end{center}
\end{figure}

To further quantify the inter-band flux-flux correlation from 1ES\,1215+303, we calculated the $Z$-transformed discrete cross-correlation function \citep[$Z$DCF;][]{Alexander13} between the light curves from different energy bands, as shown in Figure~\ref{fig:ZDCFs}. 
The $Z$DCF method offers a conservative, more efficient estimate of cross-band correlation in light curves, compared to the discrete cross-correlation function \citep[DCF;][]{Edelson88}. 
To search for time lags between these energy bands, we used a maximum likelihood function \citep{Alexander13}. 

The local peak time lag between the 3-day {\sl Fermi}-LAT and VERITAS light curve data obtained with this method is $t$(VERITAS)$- t$(LAT) $=8^{+11}_{-16}$ days compatible with a zero lag (a positive value indicates that the VERITAS flux is lagging behind the LAT flux).

There are no significant peaks in the ZDCFs for the optical and gamma or the radio and gamma or the optical and radio fluxes (this last one consistent with \citet{Lindfors2016}).

\subsection{Power spectral density of the \textit{Fermi}-LAT light curve} \label{sec:periodicity-psd}

\begin{figure}[ht!]
\begin{center}
\includegraphics[width=0.45\textwidth,angle=0]{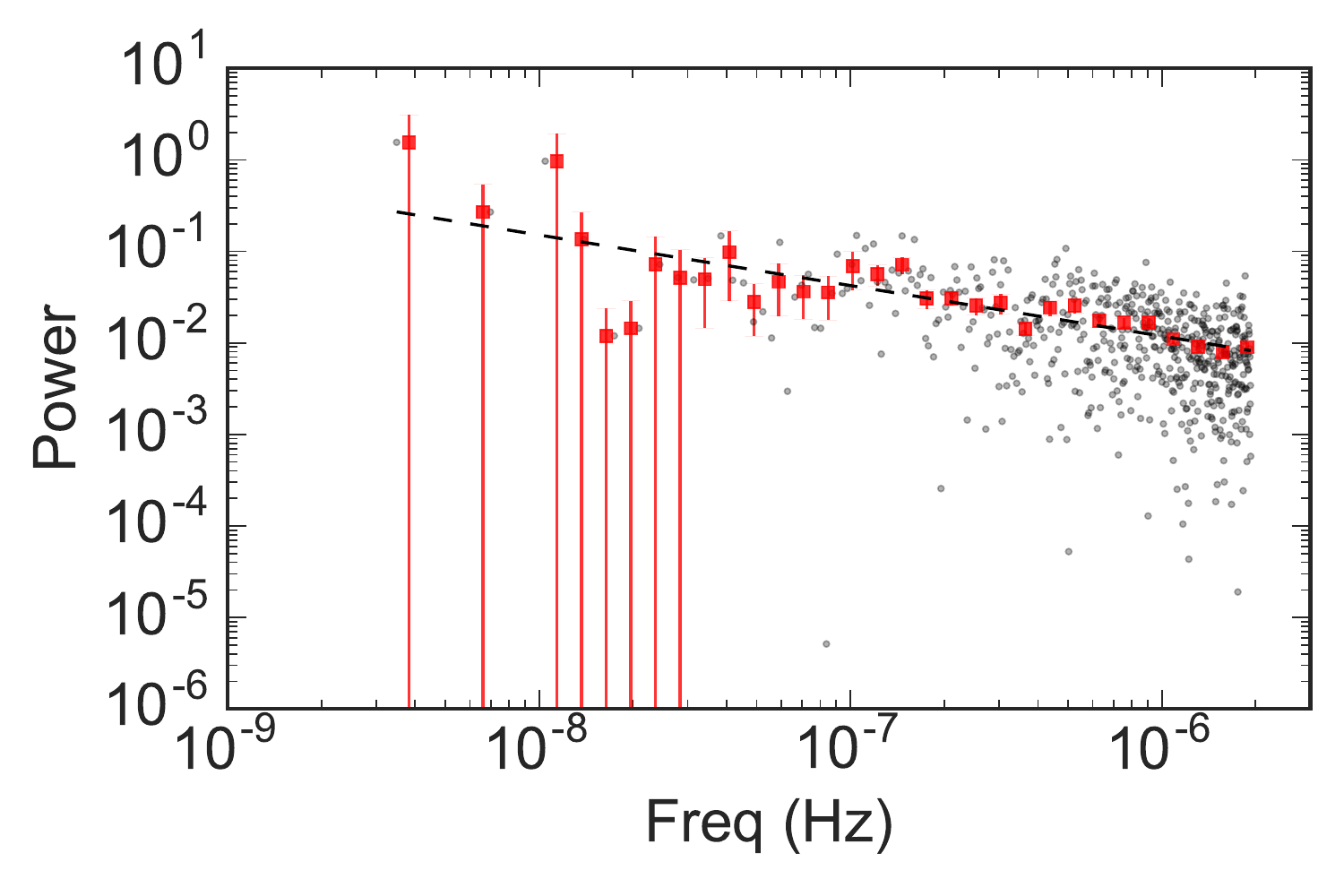}
\includegraphics[width=0.45\textwidth,angle=0]{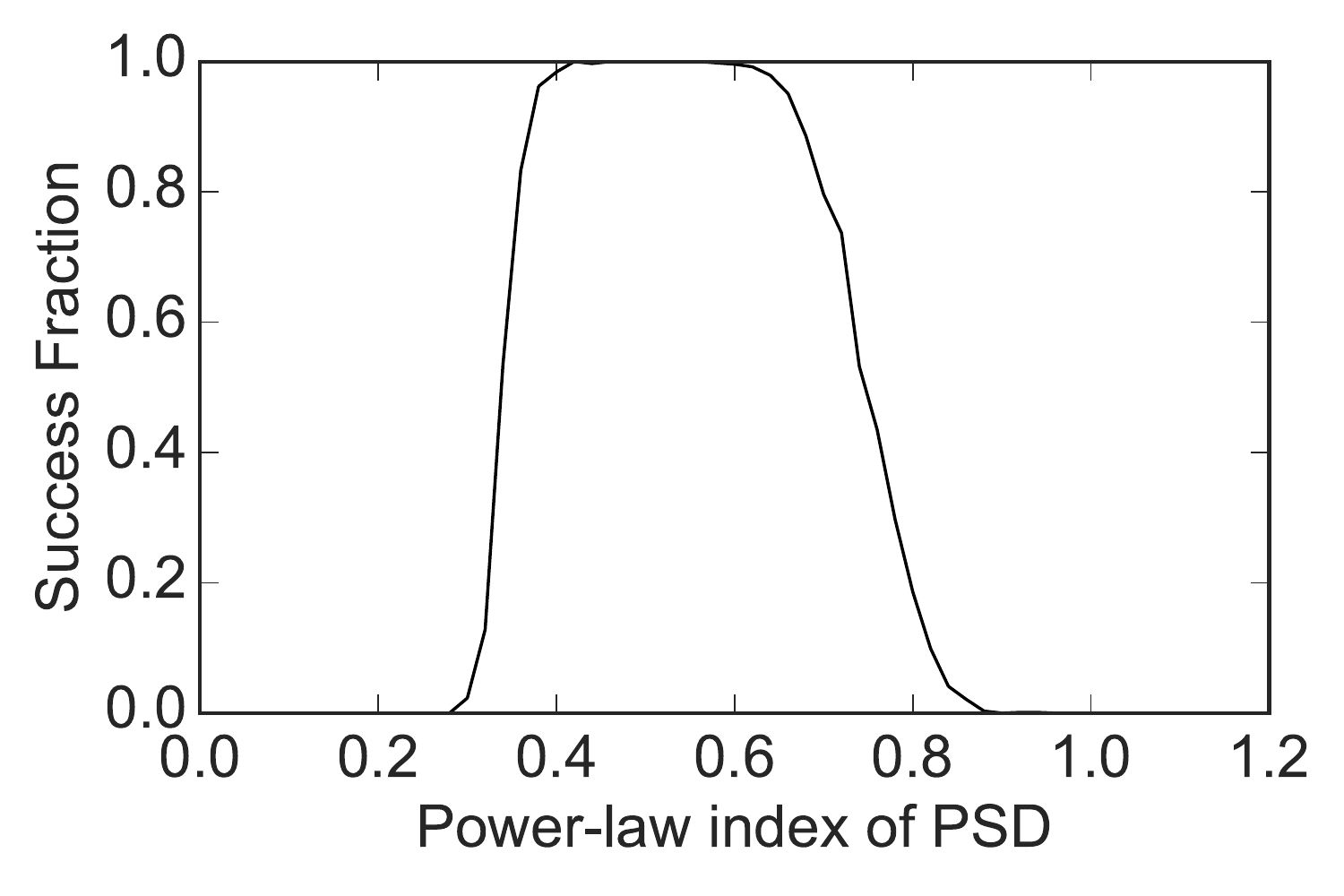}
\caption{{\it Top:} The power spectral density distribution of the 3-day-binned {\it Fermi}-LAT light curve. The gray points are the periodogram from data \citep[for details, see][]{TK95}. The red squares are the rebinned periodogram. The dashed line shows a simple power-law fit to the rebinned periodogram. 
{\it Bottom:} The ``Success Fraction'' of simulated light curves at different power-law index of the power spectral density distribution. \label{fig:PSD}}
\end{center}
\end{figure}

The source exhibits a typical power-law power spectral density (PSD) distribution, commonly observed in AGN. 
The PSD calculated \citep{TK95} from LAT data and a simple power-law fit are shown in the top panel of Figure~\ref{fig:PSD}. 
Red squares represent averages over bins with sizes that follow a geometric series of factor 1.2.

Since power-law PSDs can be distorted by power leakage from longer and shorter timescales, we calculate the ``success fraction'' (SuF) by comparing simulated light curves \citep{TK95} and the observed one, following the method described in \citet{Uttley02}. The SuF curve is shown in the bottom panel of Figure~\ref{fig:PSD}. 

The best-fit power-law index, $0.6\pm0.1$ is consistent with the relatively wide 90\% SuF range of 0.38 to 0.68. The SuF curve drops to 0 at indices of 0.3 and 0.9. 
This suggests that the PSD distribution is relatively flat compared to the typical values between 1 and 2 found in AGNs \citep[e.g.][]{Uttley02, 2014ApJ...786..143S, 2019ApJ...885...12R}. 

\subsection{ Periodicity analysis of the \textit{Fermi}-LAT and Tuorla light curves}

\begin{figure}[ht!]
\begin{center}
\includegraphics[width=0.5\textwidth,angle=0]
{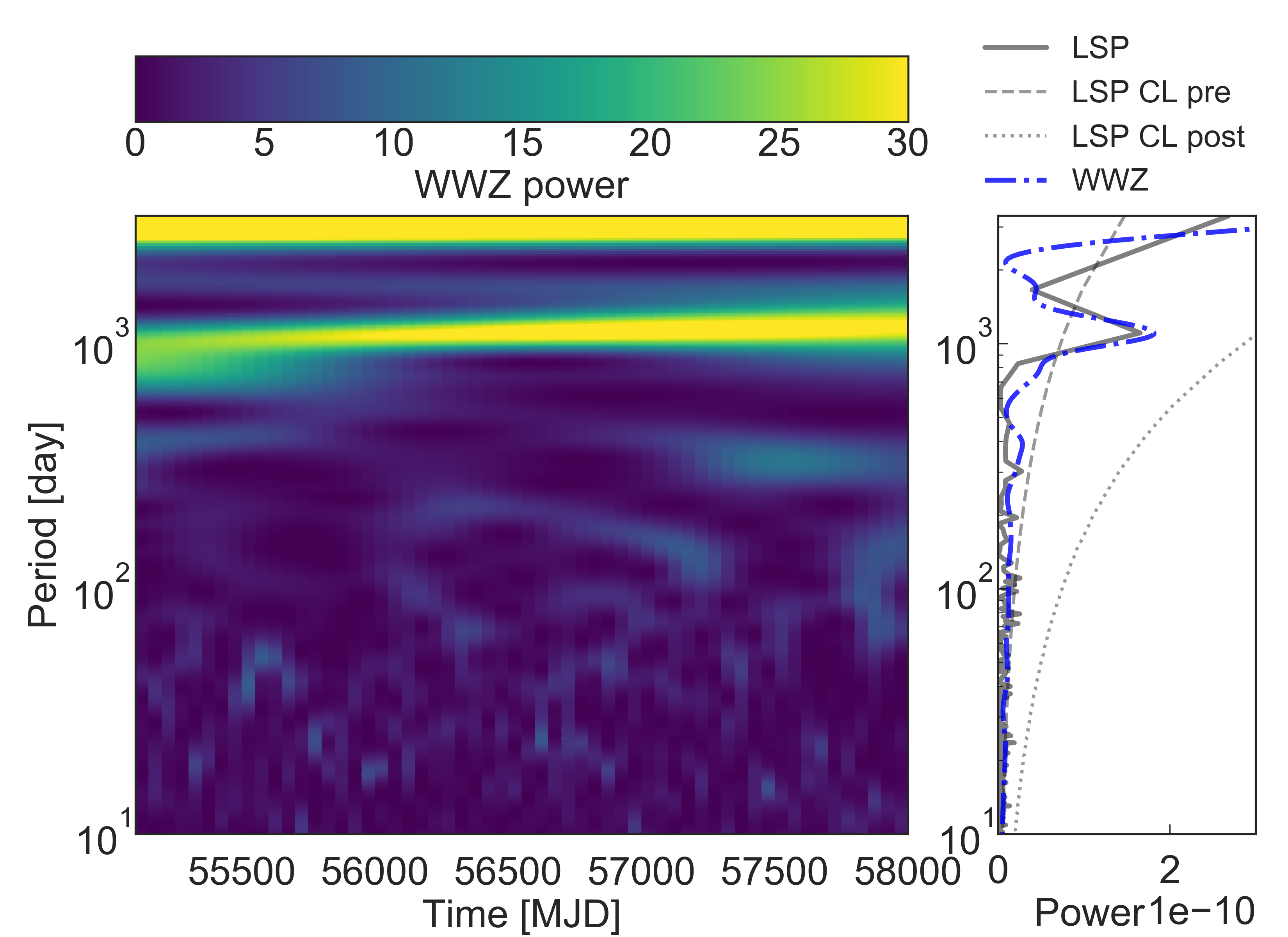}
\includegraphics[width=0.5\textwidth,angle=0]
{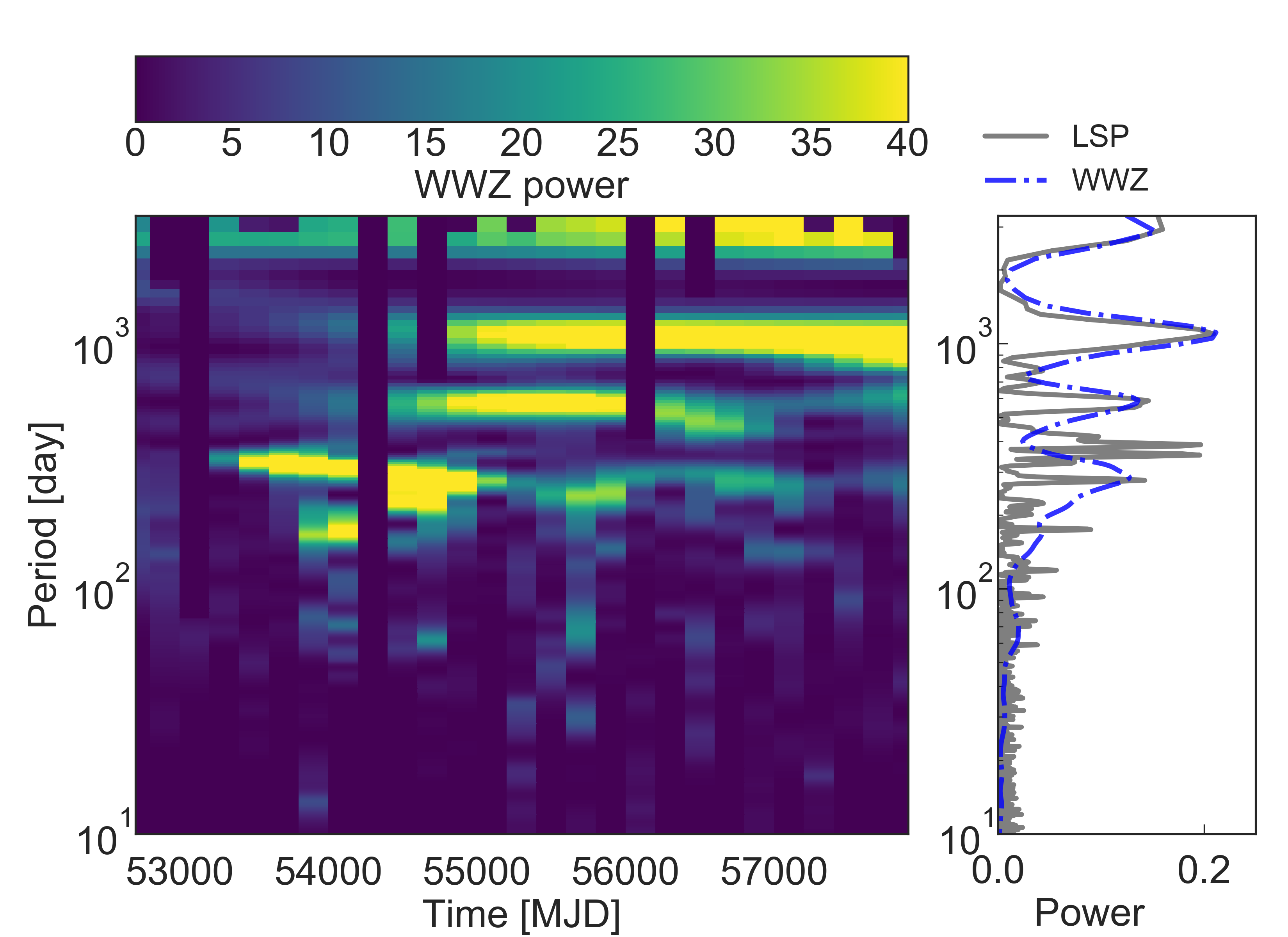}
\caption{
The scalograms from WWZ transform of the {\it Fermi}-LAT ({\it top}) and Tuorla ({\it bottom}) light curves.
The Lomb-Scargle periodogram (solid gray line) and the marginal WWZ periodogram (dash-dot blue line) are shown in the right panel of each plot. 
90\% confidence limits from a purely stochastic model with power-law PSD generated using the method of \cite{Emmanoulopoulos2013} are also shown, including (dotted gray line) and  excluding (dashed gray line) the effect of the 553 trial frequencies.
\label{fig:LSP}}
\end{center}
\end{figure}

To test for the presence of periodicity or quasi-periodic optical and gamma-ray oscillation (QPO) of the flux of 1ES\,1215+303, we calculated the weighted wavelet Z-transform (WWZ; \citealp{Foster96}) and the Lomb-Scargle periodograms \citep[LSP;][]{Scargle2} of the {\it Fermi}-LAT and Tuorla light curves, as shown in Figure~\ref{fig:LSP}. 
Both WWZ and LSP are suitable for detecting QPO in unevenly sampled light curves. 
An excess power at a $\approx\,$3-year period appears persistently in the WWZ and LSP of both the {\it Fermi}-LAT and Tuorla data throughout the observational period. Slightly lower excess power at about a half and a quarter of the $\approx\,$3-year period, and the effect of sampling gaps in the optical data are apparent in the WWZ time-frequency plot (scalogram). The {\it Fermi}-LAT LSP is noisy at shorter periods, while the periodogram (PSD) and the WWZ are much cleaner and are consistent with each other. 

The top right panel of Figure~\ref{fig:LSP} shows the PSD from the data compared with 90\% confidence limits (CL) calculated from $4.7\times10^6$ simulated light curves generated using the method of \cite{Emmanoulopoulos2013} assuming that the underlying stochastic process has a power-law PSD, and using the flux probability density function (PDF) from the right-hand panels of Figure~\ref{fig:fdist}. The dashed gray curve shows the CL for an \textit{a priori} frequency. The dotted gray curve shows the CL that includes the penalty for selecting the frequency with the largest excess \textit{a posteriori} from the 553 trial frequencies in the PSD. Assuming that the PSD is fully described by this stochastic process, it should be expected that at the 90\% CL none of the measured PSD powers exceed this dotted gray curve, and indeed none do.
Our simulations show that the apparent peaks in the LSP power at a $\approx\,$3 year period are not significant when the PSD of the underlying stochastic process and the trials factor are taken into account. The fact that the optical data show the same peak at $\approx\,$3~years does not lend credence to presence of a true QPO; this should be expected if a single stochastic process is responsible for the optical and gamma-ray light curve. 

The simulated light curves are also used to test whether the trend of linearly increasing flux found in Section~\ref{sec:increasing-flux} is inconsistent with a stationary stochastic process. We find that a linearly increasing or decreasing trend with a magnitude equal to or greater than that seen in the LAT data is present in approximately 1 in 1,000 simulations ($p=9.6\times10^{-4}$), equivalent to a significance of $\approx\,3.3\,\sigma$. The linear trend is therefore only moderately inconsistent with the stochastic modeling.

\subsection{Characterizing the flares} \label{sec:flares}

We focus our analysis on the unpublished flares, namely LAT Flare 1, LAT Flare 7 and LAT Flare 8, especially LAT Flare 7, since its peak is coincident with VERITAS Flare 7.

The decay times of {\sl Fermi} Flares 1, 7 and 8 were calculated by fitting the 1-day binned light curve to: $F(t)=F_0+F_1\times 2^{-(t-t_0)/t_{\rm{var}}}$.
The size, $R$ and Doppler factor, $\delta$, of the gamma-ray emitting region are related, due to causality, to the variability timescale through: $R\delta^{-1}\leq ct_{\rm{var}}/(1+z)$. 
The values found are shown in Table \ref{tab:tvars}.

\begin{deluxetable}{l|ccc}[ht!]
\tablecaption{The half times for the LAT flares. \label{tab:tvars}}
\tabletypesize{\scriptsize}
\tablehead{
\colhead{Flare} & \colhead{MJD}  & \colhead{$t_{\rm{var}}$ UL(90\%)}	& \colhead{$R\delta^{-1}\leq$}\\
\colhead{}           & \colhead{} & \colhead{days}
& \colhead{$10^{15}$ cm}}
\startdata
  Flare 1 & 54751 & 1.57 & 3.6\\
  {\bf Flare 7} & {\bf 57844\tablenotemark{a}} & {\bf 0.90} & {\bf 2.1}  \\
  Flare 8 & 57855 & 1.24 & 2.8\\
\enddata
\tablenotetext{a}{Coincident with a VHE flare.}
\end{deluxetable}

A similar fit was performed to the nightly VHE gamma-ray light curve around the time of Flare 7 on 2017 April 01. 
The exponential decay time was relatively well constrained at $10\,\pm \,2$ days. 
While the rise time is less constrained by the fit, we estimate the doubling time to be $<4$ days based on an upper limit measured eight days before the flare. 

From the SED modeling that we performed (as described in Section~\ref{sec:modeling}), the Doppler factor for the blob is estimated to be $\delta=25$. 
From fundamental-plane-derived velocity dispersion, \cite{Woo02} estimated the SMBH mass of the source to be $1.3\times10^{8} M_\odot$, which corresponds to a Schwarzschild radius of $R_s \sim 3.9\times 10^{11}$\,m. 
Therefore, the strongest constraint on the size of the emitting region based on the observed fastest gamma-ray variability (shown in Table \ref{tab:tvars}) is $R\leq 1350 \, R_\text{S}$. \\

\section{Spectral Analysis} \label{sec:spectrum}

\subsection{LAT long-term SED}\label{sec:longtermsed}

Three different spectral models were considered to describe the spectrum of 1ES\,1215+303 as measured by the LAT. These comprised a power-law (described in Section \ref{sec:VERITAS}), a log-parabola and a power-law sub-exponential cutoff model. 
For the combined dataset, the curved models were found to be preferred over the power-law model.

For the individual spectral data points plotted on the SED we used only the power-law model, since for these small data sets we found no preference for curved models. The data were analyzed in energy bands with the spectral parameters of all other sources in the model file being frozen to those values found in the global power-law analysis.

The power-law (PL) model, $dN/dE=N_0(E/E_0)^{-\Gamma}$, yields an integral flux of $(7.7\pm 0.2)\times  10^{-8}$~photons~cm$^{-2}\,$s$^{-1}$ with a significance of $\approx 129.1\,\sigma$ and a photon index, $\Gamma$, of $1.92\,\pm\,0.01$ at a decorrelation energy, $E_0$ of 1.36\,GeV.
The log-parabola (LP) model fit, $dN/dE=N_0(E/E_b)^{-(\alpha+\beta \log(E/E_b))}$, where $N_0$ is the normalization and $\alpha$ and $\beta$ are the spectral parameters at energy $E_b$, provided an integral flux of $(6.9\pm 0.2)\times 10^{-8}$~photons~cm$^{-2}\,$s$^{-1}$ with a significance of $\approx 129.3\,\sigma$, a spectral slope, $\alpha$, of 1.86\,$\pm$\,0.01 and a curvature parameter $\beta$, of 0.039\,$\pm$\,0.006 at the break energy, $E_b$, of 1\,GeV. 
From a power-law sub-exponential cutoff (plSECO) model, $dN/dE=N_0(E/E_0)^{-\gamma_1}e^{-(E/E_c)^{\gamma_2}}$, an integral flux of $(7.7\pm 0.2)\times 10^{-8}$~photons~cm$^{-2}\,$s$^{-1}$ was obtained with a significance of $\approx 129.3\,\sigma$, a lower-energy photon index, $\gamma_1$, of 1.74\,$\pm$\,0.03, a cutoff energy, $E_c$, of 22\,GeV, an exponent $\gamma_2$ of 0.40\,$\pm$\,0.06, and decorrelation energy, $E_0$, of 1.36\,GeV. 
Since the PL and the LP and also the PL and the plSECO are nested models, we use a likelihood ratio test to compare them, $TS_{\rm{curved}}=2(\log{L}_{\rm{curved}}-\log{L}_{\rm{PL}})$, where $L$ is the maximum likelihood of the fit. 
LP and plSECO are not nested, therefore, we do not compare them. 
We find that the LP is preferred over the PL model with a significance of $7.2\,\sigma$ while the plSECO is preferred over the PL with a significance of $7.5\,\sigma$. 
These results indicate a preference for curved models \citep[][]{4lac2019}, which could be an indicator of internal curvature at the source, even before entering the VHE range where the extragalactic background light (EBL) absorption has a considerable impact on the VHE flux.

\begin{figure}[ht!]
\begin{center}
\includegraphics[width=.45\textwidth,angle=0]{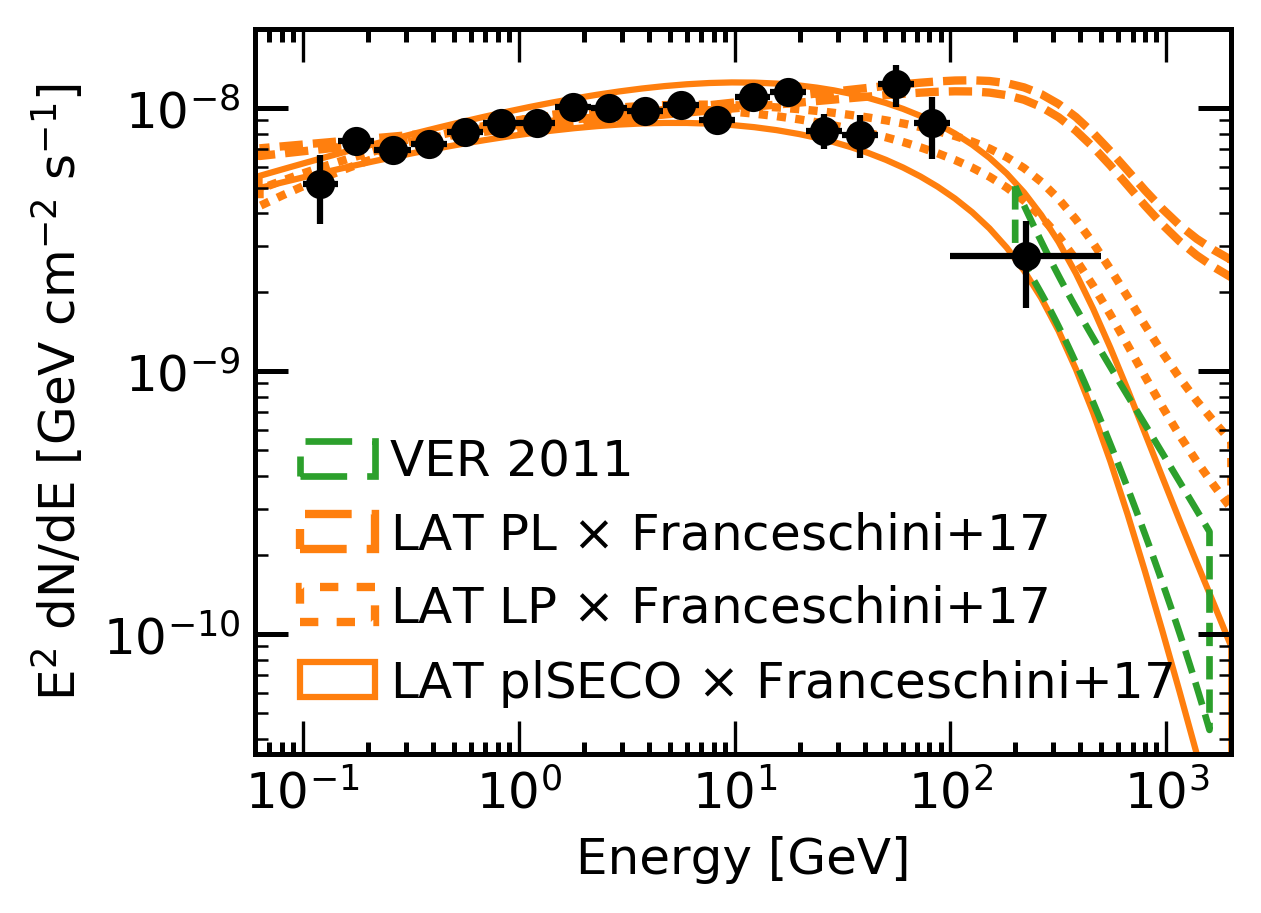}
\caption{SED of the entire {\it Fermi}-LAT data set (2008-08-04\,-\,2017-09-05). The data were analyzed with three different spectral models as described in the text: power-law (dashed), log-parabola (dotted) and power-law sub-exponential cut off (solid line). To visualize the connection with the VHE data, the VERITAS butterfly for the data from 2011 \citep{Aliu13} was added. The LAT butterflies have been extrapolated to VHE energies and the effects of the EBL included \citep{Franceschini17}. See details of the {\sl Fermi}-LAT data in Table \ref{tab:LATsed9y} in Appendix \ref{app:LATsed9y}.\label{fig:9yrsLATsed}}
\end{center}
\end{figure}

The three different fit models are shown in Figure \ref{fig:9yrsLATsed}, where the EBL absorption was taken into account by calculating interpolated values from the model of \cite{Franceschini17} (and Corrigendum \cite{Franceschini2018}) at $z=0.131$\footnote{Only the main paper is cited later in this work.}.  
The VERITAS spectrum for the 2011 data \citep{Aliu13} is shown for visualization.  
We found that the HE and VHE data are connected very smoothly. We note that the LAT spectra of the curved models are in better agreement with the VERITAS data (in this case corresponding to an average quiescent state) than the power-law spectral model.

\subsection{GeV-TeV SEDs}\label{sec:seds}

\begin{figure*}[ht!]
\begin{center}
\includegraphics[width=\textwidth,angle=0]{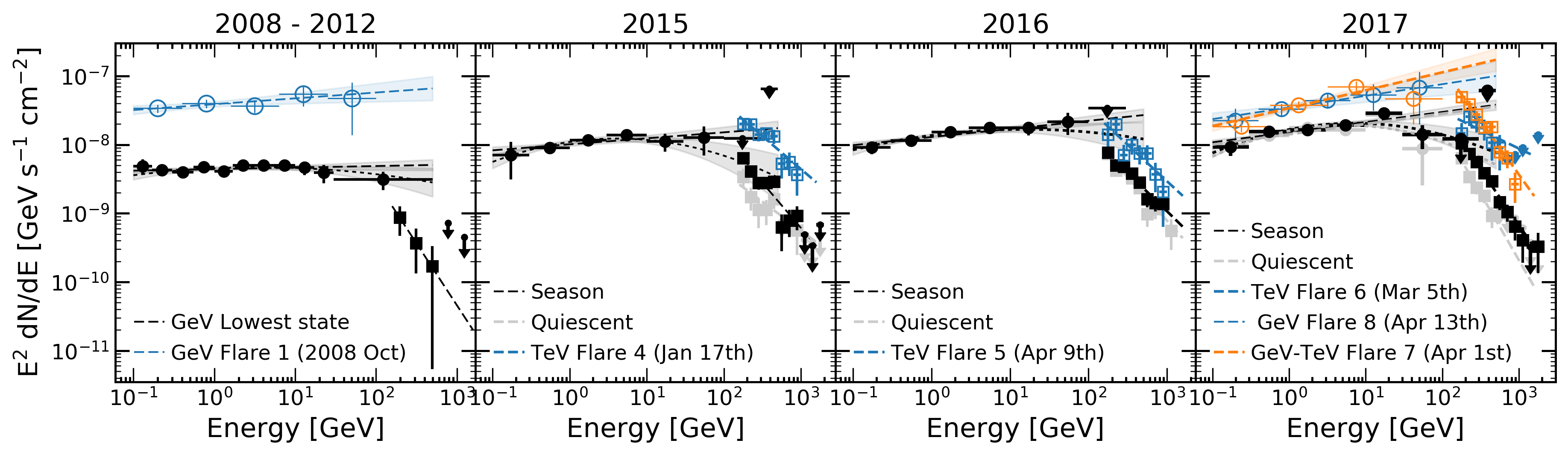}
\caption{SEDs for the LAT and VERITAS data, including flares that have not previously been analyzed. Round points correspond to the {\it Fermi}-LAT data, while the squares correspond to VERITAS. Data and butterflies for the flaring states are shown in blue and orange. Data in the quiescent state are shown in gray. From 2015 to 2017, the black data points correspond to the total data sets for each season. Power-law and log-parabola butterflies are shown for the black spectra. Only power-law butterflies are shown for the flaring states. Non-coincident GeV-TeV flare SEDs are shown in blue, while the orange SED represents Flare 7.
\label{fig:seds}}
\end{center}
\end{figure*}

The LAT-VERITAS SEDs for the unpublished VHE data are shown in Figure~\ref{fig:seds}. 
In 2008, the brightest flare (Flare 1) at GeV energies was detected.
There are, however, no corresponding VERITAS data, since 1ES\,1215+303 observations did not start until 2008 December. 

\begin{deluxetable*}{l|cccccc|cccccc}
\setlength\tabcolsep{1.5pt}
\tablecaption{Gamma-ray contemporaneous spectral analysis. \label{tab:spectra}}
\tabletypesize{\scriptsize}
\tablehead{
\colhead{Year} &\multicolumn{6}{c}{VERITAS}                                                                                     & \multicolumn{6}{c}{{\it{Fermi}}-LAT}                                            \\
\colhead{}     &\multicolumn{2}{c}{All}             & \multicolumn{2}{c}{Flare}           & \multicolumn{2}{c}{Non-flare}       & \multicolumn{2}{c}{All}             & \multicolumn{2}{c}{Flare}           & \multicolumn{2}{c}{Non-flare}     \\
\colhead{}     &\colhead{$\Gamma$} & \colhead{Flux} & \colhead{$\Gamma$} & \colhead{Flux} & \colhead{$\Gamma$} & \colhead{Flux} & \colhead{$\Gamma$} & \colhead{Flux} & \colhead{$\Gamma$} & \colhead{Flux} & \colhead{$\Gamma$} & \colhead{Flux}
}
\startdata
2015          &$3.32\pm0.18$ & $0.61\pm0.14$ & $2.96\pm0.18$ & $4.36\pm0.99$ & $2.84\pm0.39$ & $0.56\pm0.27$ & $1.91\pm 0.04$ & $5.6\pm 0.3$ & -- & -- & -- & -- \\
2016           & $3.12\pm0.13$ & $1.07\pm0.16$ & $3.06\pm0.28$ & $2.93\pm0.89$     & $3.27\pm0.14$ & $0.78\pm0.13$ & $1.88\pm 0.03$ & $7.2\pm 0.3$ & -- & -- & -- & -- \\
2017           & $3.62\pm0.10$ & $0.013\pm0.003 ^\dagger$        & $3.56\pm0.13$ &  $0.073\pm0.023 ^\dagger$           & $3.94\pm0.32$           & $0.21 \pm 0.10$        & $1.85\pm 0.03$ & $8.6\pm 0.3$ & $1.61\pm 0.32$ & $52.3\pm 25.1$ &  $1.85\pm 0.04$ & $8.0\pm 0.5$ \\
\enddata
\tablecomments{The flux value for VERITAS is the normalization (N$_0$) for the differential flux ($dN/dE$) at energy of 1 TeV in units of 10$^{-12}$TeV$^{-1}$cm$^{-2}$s$^{-1}$. The flux value for {\it Fermi}-LAT is the normalization (N$_0$) at the decorrelation energy of 1.36\,GeV in units of 10$^{-12}$MeV$^{-1}$cm$^{-2}$s$^{-1}$. $^\dagger\,$Normalization at 3 TeV. }
\end{deluxetable*}

The first panel on the left shows Flare 1 and the low state SED, as defined in Table~\ref{tab:LATobs}.
The 2011 VERITAS butterfly \citep{Aliu13} is shown since this season belongs to the GeV low state (see Section~\ref{sec:increasing-flux} for the Bayesian Blocks analysis that was used to define the various emission states of 1ES\,1215+303).

During Flare 4, at VHE, in 2015, there were approximately 40 minutes of simultaneous observations between the LAT and VERITAS; and during Flare 5, also at VHE, in 2016, there were approximately 80 minutes of simultaneous observations between the LAT and VERITAS. No significant HE emission was detected during these simultaneous observations; and no elevated flux was observed in the LAT data for these days. 
VERITAS detected another flare on 2017 March 05, Flare 6,  at a time during which 1ES~1215+303 was not in the LAT FoV. 1ES~1215+303 had been in the FoV of the LAT approximately 2.5 hours before VERITAS started observations, and re-entered the LAT FoV approximately 1 hour after VERITAS finished observing this source during that night.
No evidence for an elevated flux was found when the LAT data for this day were analyzed.
In 2017, two flares were measured by the LAT with peaks on April 01 and 13 (Flares 7 and 8, respectively; refer to Table \ref{tab:LATobs} for the duration of these flares). 
LAT Flare 7 had a VHE counterpart (orange), while VERITAS was not observing at the time of Flare 8 at GeV energies (blue). 
The details of their spectra can be found in Table~\ref{tab:spectra}.\\

\section{Multifrequency radio-to-TeV SED modeling} \label{sec:modeling}

The large multiwavelength dataset described in this paper allows us to build broadband SEDs for different periods and states of activity of 1ES~1215+303. 
In this section, three activity states that have not been examined in previous works are studied: a low, steady state corresponding to the lowest observed \textit{Fermi}-LAT activity as defined by the Bayesian Block method, the 2017 April 01 GeV-TeV Flare 7, and the subsequent post-flare state from 2017 April 15 to 23.

These three states are modeled using the ``blob-in-jet" (Bjet) radiative code from \cite{Hervet15}. 
Given the low apparent jet speeds reported in Section \ref{sec:vlba}, we consider the main emission zone as a continuous high-energy particle flow passing through a stationary shock in the jet. This local plasma flow is identified as a compact spherical blob flow moving at a significant Lorentz factor close to the line of sight.
We assume that this blob is filled by an electron (or electron/positron) population in an isotropic magnetic field.
We consider a particle energy distribution which, as a result of injection and cooling, follows a broken power-law function as
\begin{equation}
N_e(\gamma) = \left\lbrace
     \begin{array}{ll}
         N_{e}^{(1)} \gamma^{-n_1} & \mathrm{for}~ \gamma_{\mathrm{min}} \leqslant \gamma \leqslant \gamma_{\mathrm{brk}} \\
         N_{e}^{(2)} \gamma^{-n_2} & \mathrm{for}~ \gamma_{\mathrm{brk}} \leqslant \gamma \leqslant \gamma_{\mathrm{max}}
     \end{array}
     \right. ,
\end{equation}
with $N_{e}^{(2)} = N_{e}^{(1)}  \gamma_{\rm{brk}}^{(n_2-n_1)}$, and $N_{e}^{(1)}$ the particle density factor set as $N_{e}^{(1)} =N_e(1)$.

This blob is moving through a conical leptonic plasma jet having a larger radius and a lower flow Lorentz factor. 
The jet is discretized logarithmically into 50 conical slices along its propagation axis. For the sake of simplicity, each slice has its particle density spectrum considered as a simple power-law function.
Both the blob and the jet are radiating in synchrotron and SSC emission. 
We include the effects of the absorption by the EBL following the model of \cite{Franceschini17}.
We model the data via a ``fit by eye" process, because the use of a minimization algorithm is very challenging for SSC models due to the strong degeneracies that exist between parameters. Furthermore, it becomes extremely difficult when we have multiple emission zones such as is considered here. Hence the proposed model solutions cannot be considered as the statistically best solutions but are consistent with our assumptions about the underlying emission scenario. The reduced $\chi^2$ of the fits shown in the following section is for informational purposes only.

\subsection{Low state of 1ES 1215+303}

\begin{figure*}[ht!]
	\begin{minipage}[b]{0.5\linewidth}
   		\centering \includegraphics[width=9.5cm]{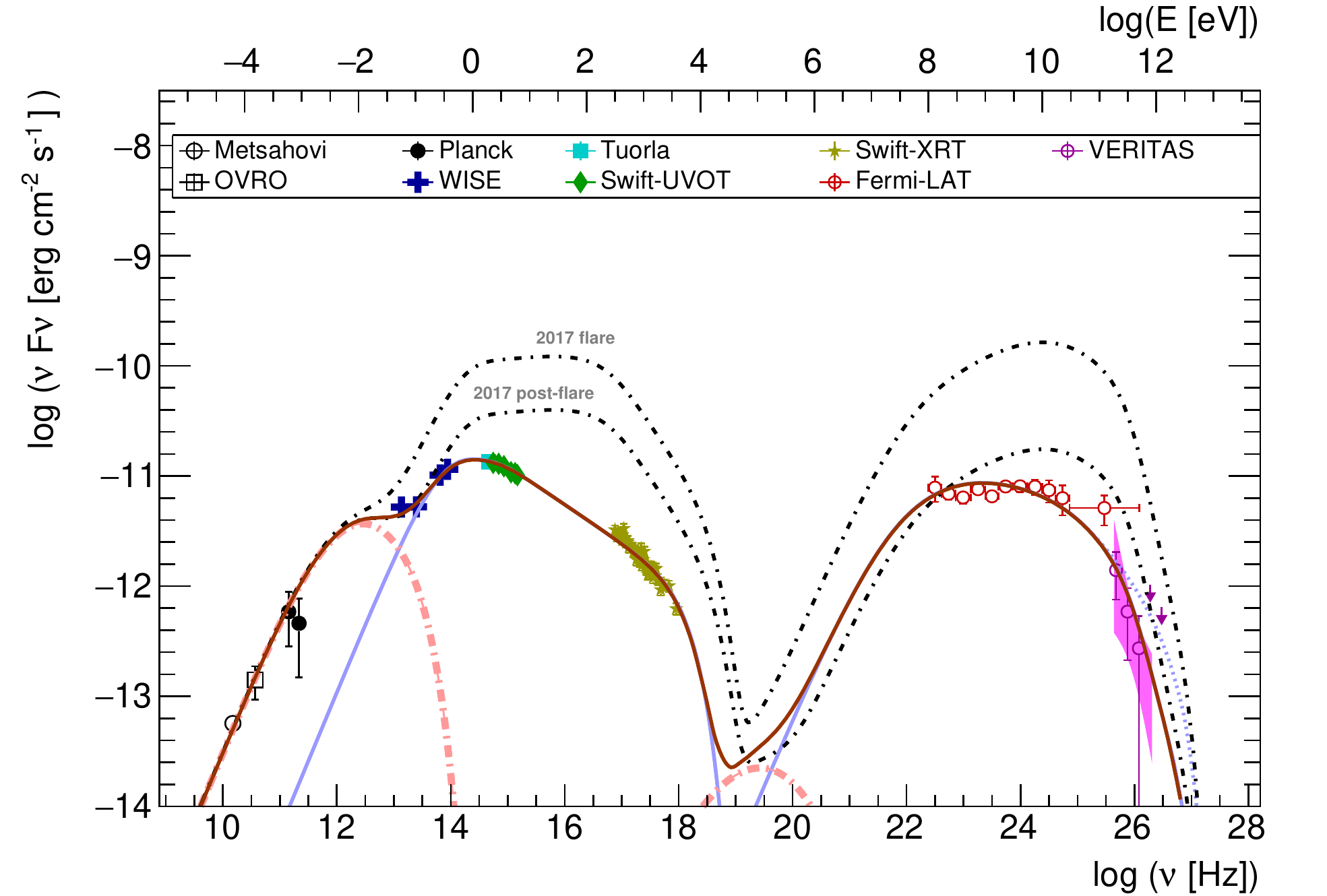}
	\end{minipage}\hfill
		\begin{minipage}[b]{0.5\linewidth}
      \centering \includegraphics[width=9.5cm]{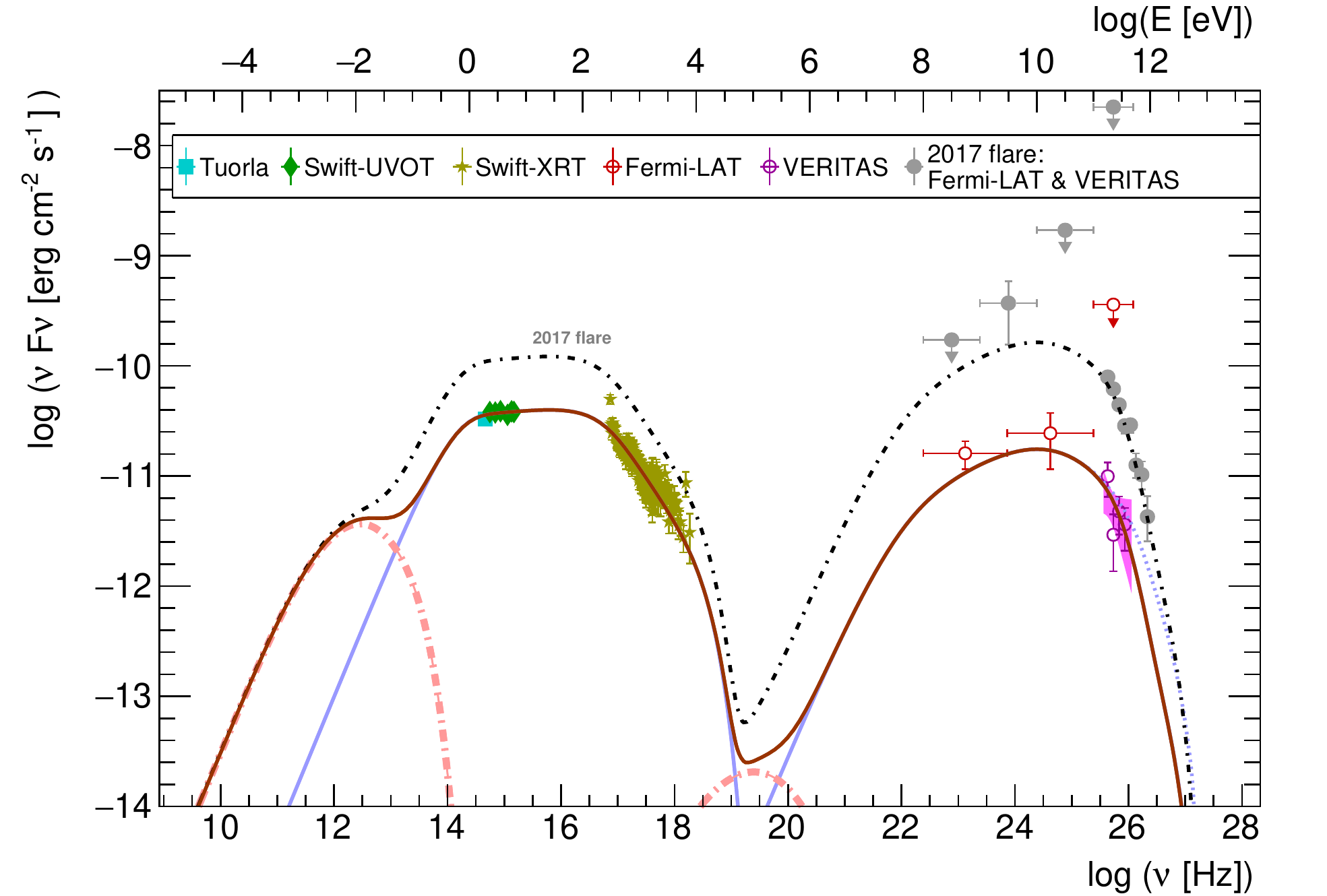}
	\end{minipage}\hfill
\caption{Multiwavelength SEDs and models of the source low state (\textit{Left}), 2017 flare and 2017 post-flare (\textit{Right}). Plain blue lines are the blob synchrotron and SSC contributions, dot-dashed pink lines are the jet synchrotron and SSC emission, the blue dotted line is the intrinsic SSC emission without EBL absorption, and the thick brown and thick black dot-dashed lines are the sums of all components.
\label{fig:SEDmodel_lowstate}}
\end{figure*}

The time period corresponding to the low state of the source was defined using the results of the Bayesian block method that was applied to the \textit{Fermi}-LAT lightcurve (see Fig.~\ref{fig:gev-tev-lcs}). 
Two periods between 2008 and 2012 can be considered as the lowest activity state: 2008 November 17 -- 2010 August 12 (MJD 54787--55421) and 2011 April 15 -- 2012 April 10 (MJD 55666--56027). 
The multiwavelength lightcurves do not show any evidence for an outburst occurring at other wavelengths during these time periods either.
Such a long accumulated time of 33 months of low state allows us to have a very well defined {\it{Fermi}}-LAT spectrum, as well as a well-sampled multiwavelength SED at lower energies. 
Indeed, data from the \textit{Planck PCCS2} catalog \citep{Planck16} and the AllWISE Multiepoch Photometry Database\footnote{\url{http://wise2.ipac.caltech.edu/docs/release/allwise/}} were taken during our defined periods, increasing the broadband coverage.
The resulting SED with the favored associated radiative model is presented in Figure~\ref{fig:SEDmodel_lowstate}, and the model parameters are shown in Table~\ref{tab::Params_model}. The favored model has a $\chi^2/\rm{d.o.f.} = 364./49 = 7.4$ (considering the blob and jet model parameters). The fit quality is strongly impacted by the extremely small uncertainties of the averaged \textit{WISE} data. Without taking into account \textit{WISE}, we have $\chi^2/\rm{d.o.f.} = 106./45 = 2.4$.

\subsubsection{Compact blob}

The multiwavelength SED from the IR to gamma ray is assumed to be emitted from a compact emission zone, referred to above as the ``blob."
The SED shows two clear bumps, one peaking in the IR-optical range considered as synchrotron emission and one peaking at high energy considered as being dominated by SSC emission. 
The apparent contradiction with this observed low frequency synchrotron peak and the HBL classification of the source is further discussed in Section~\ref{sec:discussion}.

Neither the thermal signature of accretion disk radiation nor a sharp peak at high energy, which would indicate the presence of the external inverse-Compton (EIC) process on the nucleus thermal radiation field, is detected. 
We therefore consider this process to be negligible, as is often the case for HBL sources.

The wide gap in energy of about ten orders of magnitude between the synchrotron and SSC peaks implies a very low internal $\gamma-\gamma$ opacity to reach the observed energies of $E > 100$\,GeV. 
A satisfactory solution is found by considering a high Doppler factor value of $\delta = 25$, associated with the maximum theoretical angle to the line of sight $\theta \simeq 2^{\circ}$.
As described in Section~\ref{sec:flares}, the radius of the emitting region is constrained by taking into account the fastest observed variability of $t_{\mathrm{var}} = 0.9$ day. 
Given the Doppler factor considered, this sets an upper limit to the radius of $R \leq 5.2 \times 10^{16}$ cm.

The minimal energy of the radiative electrons is set at the relatively high value of $\gamma_{\mathrm{min}} = 4.7\times 10^{3}$. 
While not exceptional in blazar radiative models, such a high $\gamma_{\mathrm{min}}$ is often specifically used to describe extreme HBLs \citep[e.g][]{Aliu14,Archer18}.
The blob is matter-dominated with an equipartition ratio between the magnetic field energy density $U_B$ and the particle energy density $U_e$ of $U_B/U_e = 1.6 \times 10^{-2}$.

\subsubsection{Radio jet}
\label{sec:Modeling-radio_jet}

The \textit{WISE} SED shows a clear luminosity excess in its lowest energy band W4 compared to the other ones, which follow a hard photon index power-law spectrum, as expected for the optically thick blob synchrotron emission.

This excess can be associated with broader jet emission, dominating the low-energy part of the SED from radio to far infrared. 
Although not often modeled, this jet signature is a relatively common HBL feature \citep[e.g.][]{Katarzynski01, Archer18}.

With 9 free parameters and only one obvious spectral signature in the radio to far IR, the jet parameters are naturally degenerate. 
In order to have parameters as physically consistent as possible while keeping a good fit to the data, we constrain several other parameters in addition to the density and Doppler factor that are discussed above. 
We consider an identical spectral slope for the injected particle spectrum between the blob and the jet and we also assume that the jet is in equipartition.

The apparent opening angle of the 15.3\,GHz radio-jet was measured as $\alpha_\mathrm{app} = 13.8^{\circ} \pm 0.1^{\circ}$ by \cite{Pushkarev17} {\em via} a stacking of the multiple observations of the VLBA referenced in the MOJAVE database. 
This value confirms the previous measurement of $\alpha_\mathrm{app} = 14^{\circ}$ by \cite{Hervet16}, which was derived from the same database but based on the evolution of the referenced radio-component sizes.
The fact that these two measurements are similar indicates that the jet does not significantly change its direction with the line of sight over time, and that the radio components occupy the full jet cross-section.

From the observed jet apparent opening angle and the angle with the line of sight set at $\theta = 2^{\circ}$, we can deduce the intrinsic jet opening angle used for the model {\em via} the relation
$\alpha = \alpha_\mathrm{app} \sin(\theta)$, which leads to $\alpha/2 = 0.24^{\circ}$.

\begin{deluxetable}{ccc}
\tablecaption{Model parameters used for the multiwavelength low state.\label{tab::Params_model}}
\tabletypesize{\scriptsize}
    \tablehead{
    \colhead{Parameter} & \colhead{Value} & \colhead{Unit}\\ \hline
    \colhead{$\theta$} & \colhead{$2.0$} & \colhead{($^{\circ}$)}\\ \hline
    \colhead{Blob} &  &
    }
    \startdata
    $\delta$ & $25$ & $-$\\ 
    $N_{e}^{(1)}$ & $1.8\times 10^{6}$ & cm$^{-3}$\\ 
    $n_1$ & $2.82$ & $-$\\
    $n_2$ & $3.7$  & $-$\\
    $\gamma_{\mathrm{min}}$ & $4.7\times 10^{3}$ & $-$\\
    $\gamma_{\mathrm{max}}$ & $7.0\times 10^{5}$ & $-$\\
    $\gamma_{\mathrm{brk}}$ & $1.5\times 10^{4}$ & $-$\\
    $B$ & $2.35\times 10^{-2}$ & G\\
    $R$ & $5.1\times 10^{16}$ & cm\\
    \hline
    Jet\\
    \hline
    $\delta$ & $15$ & $-$\\ 
    $N_{e}^{(1)}$ & $1.3\times 10^{4}$ & cm$^{-3}$\\ 
    $n$ & $2.82$ & $-$\\
    $\gamma_{\mathrm{min}}$ & $9.0\times 10^{2}$ & $-$\\
    $\gamma_{\mathrm{max}}$ & $3.5\times 10^{3}$ & $-$\\
    $B_1$ & $3.5\times 10^{-2}$ & G\\
    $R_1$ & $1.0\times 10^{17}$ & cm\\
    $L$\tablenotemark{*} & $1.0\times 10^{2}$ & pc\\
    $\alpha/2$\tablenotemark{*} & $2.4\times 10^{-1}$ & $^{\circ}$
\enddata
\tablenotetext{*}{\textit{ Host galaxy frame.}}
\end{deluxetable}

\subsection{2017 April flare and post-flare}

\begin{deluxetable}{ccc}
\tablecaption{Model parameters used for the multiwavelength 2017 April 01 flare and post-flare states.\label{tab::Params_model_flare}}
\tabletypesize{\scriptsize}
\tablehead{
    \colhead{Parameter} & \colhead{Value} & \colhead{Unit}\\ \hline
    \colhead{$\theta$} & \colhead{$2.0$} & \colhead{($^{\circ}$)}\\   \hline
    \colhead{Blob} &  &
    }
    \startdata
    $\delta$ & $25$ & $-$\\ 
    $N_{e}^{(1)}$ (flare) & $\boldsymbol{5.5\times 10^{6}}$ & cm$^{-3}$\\ 
    $N_{e}^{(1)}$ (post-flare) & $\boldsymbol{1.8\times 10^{6}}$ & cm$^{-3}$\\ 
    $n_1$ & $2.9$ & $-$\\
    $n_2$ & $4.5$  & $-$\\
    $\gamma_{\mathrm{min}}$ & $4.7\times 10^{3}$ & $-$\\
    $\gamma_{\mathrm{max}}$ & $7.0\times 10^{5}$ & $-$\\
    $\gamma_{\mathrm{brk}}$ & $9.0\times 10^{4}$ & $-$\\
    $B$ & $5.2\times 10^{-2}$ & G\\
    $R$ & $5.1\times 10^{16}$ & cm
    \enddata
\end{deluxetable}

On 2017 April 01 (MJD 57844), VERITAS detected its second brightest flare from 1ES\,1215+303 (referred to as Flare 7). 
This strong gamma-ray activity was simultaneously detected by \textit{Fermi}-LAT and was followed by a secondary \textit{Fermi}-LAT outburst 10 days later which we call {\sl Fermi} Flare 8 (see Fig.~\ref{fig:gev-tev-lcs}). 
Unfortunately 1ES\,1215+303 was not being monitored at any other energies during this time, which prevents us from being able to derive any accurate emission scenario for this April 01 event.

From 2017 April 15 to 23 (MJD 57858--57866), the source was monitored at many wavelengths and showed historically high fluxes in the optical, UV, and X-ray bands (see Fig.~\ref{fig:all-lightcurves}).
It is plausible then that the emission zone responsible for the {\sl Fermi} gamma-ray Flares 7 and 8 was still in its cooling phase during this period.

Given the many multiwavelength observations available during this post-flare period, we can attempt to derive realistic physical parameters describing the data. 
As is shown in Figure~\ref{fig:SEDmodel_lowstate} and Table~\ref{tab::Params_model_flare}, a particle density decrease of a factor 3 in the emission zone is enough to move from the flare to the post-flare state. 
Such a decrease matches an interpretation of a flare from a jet overdensity crossing a standing shock.

The radio jet is assumed to keep a roughly steady flux between all of the states studied. 
The jet model used for the low state is kept for the 2017 flare/post-flare, and plays only a very minor role in the total radiative output.

We considered the same emission zone for all of the SEDs modeled, with a constant plasma flow speed (same Doppler factor and size). 
The low- and high-state SEDs can be well represented by changing the particle spectrum and the magnetic field parameters. 
The substantial changes introduced between the low and the high states are an increase of the magnetic field $B$ ($ \times 2.2$), an increase of the particle spectral break energy $\gamma_\mathrm{brk}$ ($\times 6.0$), and a softening of the particle index after the break $n_2$ ($\times 1.2$). 
Interpretations of these changes are discussed in Section~\ref{sec:discussion}. The fit quality of the flare and post-flare states is $\chi^2_{\rm flare}/\rm{d.o.f.} = 10.5/0$ and $\chi^2_{\rm p-flare}/\rm{d.o.f.} = 385/184 = 2.1$ respectively.

\section{Discussion} \label{sec:discussion}

\subsection{Extreme shift of the synchrotron peak frequency}

In many ways 1ES\,1215+303 shows typical features of a classical HBL source: it has an FR I radio jet \citep[at the kpc scale]{Giroletti06}, with multiple stationary radio-components as can be seen from VLBI \citep{Hervet16,Piner18}, it does not show a thermal accretion disk signature in the blue-UV, nor does it exhibit strong inverse-Compton dominance in the broadband SED.

An unusual feature, however, is the dramatic change of the synchrotron bump (shape and peak frequency) between low and high activity states. 
The high state, as observed in the 2017 flare and post-flare SED, presents a synchrotron peak between the UV and soft X-rays, typical of HBLs. Due to the relative flatness of the synchrotron bump it is difficult to determine the precise peak frequency value, but the favored post-flare model shows a synchrotron peak at $\log_{10}(\nu_{\rm{peak}}/\rm{Hz}) = 15.75$.
The low state is characterized by a much more constrained peak frequency, $\log_{10}(\nu_{\rm{peak}}/\rm{Hz})$, of 14.49\,$^{+0.17}_{-0.54}$ from the model, with boundaries from the IR and optical data (consistent with \citet{Nilsson2018}'s results, based on the Roma-BZCAT Multi-frequency Catalogue up to 2012). 
Thus, if only this low state were considered, this source would be classified as an IBL.

\begin{figure}[ht!]
\begin{center}
\includegraphics[width=.49\textwidth,angle=0]{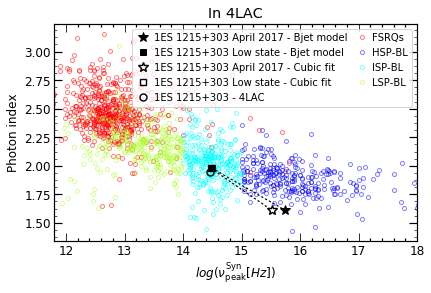}
\caption{Photon index versus the logarithm of the frequency of the synchrotron peak. Color markers represent classifications, indicated in the legend, for GeV-detected blazars as published in the 4LAC. 1ES 1215+303 shows a spectral shape characteristic of IBLs during the low state, while exhibiting HBL-like properties during the high state in April 2017. This extreme shift is observed with both the results of the blob-in-jet modeling and the cubic polynomial fit (see text for details).\label{fig:shift4LAC}}
\end{center}
\end{figure}

Fits to a cubic polynomial function were also performed on the synchrotron bump of the broadband SED; since this is the method followed in the Fourth Catalog of AGNs detected by the {\sl Fermi}-LAT \citep[4LAC;][]{4lac2019}. The results were consistent with the blob-in-jet modeling, and are illustrated in Figure \ref{fig:shift4LAC}.

Up to now, the only extreme peak-frequency shift ever observed from mid-IR to X-ray is from the IBL VER\,J0521+211 (also known as RGB\,J0521+212) with, however, a lack of optical-UV data during its flare, which prevents any reliable peak shift estimation \citep{Archambault13}.
HBLs are also subject to synchrotron peak shifts during flares but to a lesser extent, e.g., between soft/mid to hard X-rays, making a transition possible from regular to extreme HBL \citep[e.g][]{Ahnen18}.
Thus, the reported frequency shift in this study is a first for this kind of source, which further increases the diversity of behaviors observed for BL Lacs and raises many questions about the causes of such phenomena.

A critical parameter illustrating this synchrotron peak shift is the Lorentz factor break of the electron spectrum, $\gamma_{\mathrm{brk}}$, which increased by a factor of 6 between the low and high states.
Following the common broken power-law description of the particle spectrum, the $\gamma_{\mathrm{brk}}$ parameter represents the energy above which the radiative cooling is taking over from the adiabatic (or advective) cooling \citep[e.g][]{Inoue96}. 
A significant increase of $\gamma_{\mathrm{brk}}$, as suggested by the SED modeling, points towards more efficient adiabatic cooling when flaring.
In order to have a flare with more efficient non-radiative cooling, the model shown in Figure \ref{fig:SEDmodel_lowstate} requires a strong increase of the population of injected particles in addition to a local increase of the magnetic field.
Due to the degeneracy between the magnetic field and the Doppler factor in blazar SSC models, a local increase of the Doppler factor instead of the magnetic field is also a possible explanation.

The linear flux-flux correlation between the optical and the GeV gamma-ray bands highlighted in Section \ref{sec:fluxflux}, showing an index ($a=0.86$) of less than 1, is consistent with the fact that a larger variation of the synchrotron peak luminosity than the SSC one was observed in the low state and 2017 post-flare SEDs.
The exclusion of a quadratic flux-flux correlation indicates that a change in the number of radiative particles is not the major criterion explaining the common observed variability. However this could be favored for the strongest flares, such as that of 2017 April 01 (see Figure \ref{fig:SEDmodel_lowstate} and Table \ref{tab::Params_model_flare}).

\subsection{Multi-year flux increase}

The broken-line fit of the long term lightcurves is strongly favored over the linear fit for the \textit{Fermi}-LAT dataset ($5.5 \,\sigma$ level), and moderately favored for the optical dataset ($3.4 \,\sigma$ level). The times where the break occurs in both datasets are compatible within $1\,\sigma$, strengthening the case for a MWL increase of the source activity starting approximately at the time of MJD $55780 \pm 122$ ($\sim$ 2011 August).

Even though the LAT linear trend is inconsistent with the stochastic model only at the $3.3\,\sigma$ level (see Section \ref{sec:periodicity-psd}), this long term flux increase of at least 6 years is intriguing and can be caused, in theory, by multiple possible processes such as jet precession, or by an increase in the accretion rate.

The multiple radio-VLBI observations of the source presented in this work, such as the lack of non-radial motions in the jet, and the straight jet at larger scales discussed in \cite{Giroletti06} rule out any significant jet precession. 
Also, jet precession would make the jet width, from stacked radio images, broader than the measured component sizes  (Section \ref{sec:Modeling-radio_jet}).
Finally, any precession would likely lead to a long-term rise in the radio emission due to the increase of the Doppler factor. None is observed.
We thus consider that the most likely cause of this gamma-ray multi-year flux increase is related to the black hole accretion process. 

Tidal disruption events (TDEs) are often mentioned when observing multi-year-long flares of supermassive black holes. 
These should be at a particularly high rate in AGNs due to the interaction of their accretion disk/torus with nearby stars \citep{Karas07}.
It is, however, very challenging to differentiate a TDE from the natural high-amplitude variability of the accretion disk itself. 
A TDE is usually identified by its strong nuclear ionization and by a specific decreasing flux profile. 
We do not have access to these observables with the data that we have gathered, which prevents us from any relevant testing of the TDE hypothesis.

This long-term flux increase can be, however, compared to typical timescales of natural changes that occur in the accretion rate. HBLs are known to be the least powerful blazars and have been associated with a weak accretion mode known as the ``advection dominated accretion flow'' (ADAF).
In this case, the typical minimal time for jet loading from a change in the accretion is given by the free-fall timescale $\tau_{ff}$. From \cite{Manmoto96} we have
\begin{equation}\small
\tau_{ff} = 4.63 \times 10^{-5} \left(\frac{r}{1.0 \times 10^{3} r_g}\right)^{3/2} \left(\frac{M_{\mathrm{BH}}}{10 M_\odot}\right) \mathrm{days}.
\end{equation}

By considering matter loading from the outer part of the ADAF disk, at $r \sim 3.0 \times 10^{3} r_g$ \citep{Narayan96}, and the black hole mass $1.3\times 10^{8} M_\odot$ (as discussed in Section~\ref{sec:flares}), we obtain a typical timescale $\tau_{ff}$ of 8.7 years.
This timescale is similar to the long-term flux increase reported in Section \ref{sec:increasing-flux} which started around the fall of 2011.

We found evidence (significance of $4.7\,\sigma$) for a long-term spectral hardening trend accompanying the flux increase (see Sections~\ref{sec:increasing-flux} and \ref{sec:latSpectrumFlux}). 
Such a ``harder-when-brighter'' trend (at a $3.6\,\sigma$ level in the case of 1ES~1215+303) is typically observed in gamma-ray flat-spectrum radio quasars and intermediate-/low-frequency peaked BL Lacs \citep[e.g.,][]{Abdo10}. 
Similar behavior has been observed in radio galaxies and high-frequency peaked BL Lacs, most commonly in the X-ray band \citep[e.g.,][]{Brown11, Ahnen16}. 
From our SED modeling above, the GeV gamma-ray spectra during higher flux states are indeed harder than the lowest flux state, lending support to the ``harder-when-brighter'' phenomenon.

\subsection{Optical polarization}

The optical polarization fraction over the 3 years covered by the NOT observations is relatively stable, with values between 5 and 15\,$\%$. 
This relatively low blazar polarization is well within the range of small values typical of HBL sources \citep{Angelakis16}.
In the same paper, it was noted that HBLs tend to concentrate their polarization angle around preferred directions, which is also the case for 1ES\,1215+303 with small angle variations from 130$^{\circ}$ to 175$^{\circ}$.
This indicates a stable, nearly toroidal magnetic field structure at the location of the optical emission zone that we described as a compact blob.

The NOT observations provide good optical polarization coverage around the gamma-ray flare of 2017 April 01. 
During this epoch, the polarization angle reached its highest value ($173^{\circ}$), remaining above $166^{\circ}$ during the post-flare state. At the same time, the polarization fraction reached its local minimum during the post-flare state. The polarization angle local minimum of the season was $140.6^{\circ}$, varying a total of $38.4^{\circ}$ in 2017; while the polarization fraction changed between 5\% and 10.5\%.

Although this angle shift is much less dramatic than what has been observed in some blazars \citep[e.g.][]{Abdo10Natur,Marscher10,Kiehlmann16}, it follows a common behavior  associated with gamma-ray flares \citep{Blinov18,Hovatta2016}. 
The weak amplitude of the polarization angle shift could find a natural explanation in the observed almost toroidal magnetic structure and a heavily matter-dominated blob, as suggested by the modeling.

\subsection{Log-normal distribution of the optical and HE fluxes}

The preference for log-normality in the flux distributions of the LAT and Tuorla data {could be} evidence that multiplicative processes \citep{Aitchison1973} are occurring at these wavelengths, which are, as is discussed in Section \ref{sec:fluxflux}, strongly correlated over the long term, and which could also be connected due to SSC scattering. 
Several hypotheses have been discussed in the literature regarding the nature of the processes behind these observations. 
For instance, \cite{Uttley01} attribute them to large, long-time-scale energy releases in the corona, possibly due to magnetic reconnection, initiating avalanche sub-division, which is later superimposed on short-time-scale emissions of energy proportional to the original division. 
They also mention the natural appearance of these linear relationships in the mechanism proposed by \cite{lyubarskii97} due to radius-dependent mass-accretion-rate fluctuations producing variations on all time scales in the disk and corona. 
However, an interpretation based on additive processes by \cite{biteau12}, the mini-jets-in-a-jet model, predicts that skewed flux distributions (such as log-normal) could be obtained from the summation of contributions of a large number of mini-jets under specific conditions.

\section{Summary} \label{sec:summary}

In this paper we present an analysis of the observations of the HBL 1ES\,1215+303 between 2008 and 2017 from radio to VHE gamma-ray energies. We summarize our main findings below:

(i) The observations performed by {\sl Fermi}-LAT in gamma-rays and the Tuorla Observatory in optical show a clear long-term increase of flux over the ten-year period. Both datasets favor a start of this increase around August 2011 ($\approx\,$MJD $55780\pm 122$). No conclusive interpretation is found to explain such a behavior; however, the timescale of this flux increase, while limited by our dataset, is consistent with a process driven by the accretion disk. We can also reject jet precession as the cause of this behavior since precession is not in agreement with the multiple radio-VLBI observations.

(ii) We report the simultaneous coverage of the peak day of Flare 7  between the \textit{Fermi}-LAT and VERITAS instruments, occurring on the night of 2017 April 01 (MJD 57844).

(iii) An extreme shift of the synchrotron peak frequency from the low state to the 2017 flaring state of the source is observed. 
This is consistent with a higher break energy of the emitting particles in the flaring state, likely associated with a more efficient adiabatic cooling.

(iv) Three stationary radio features in the innermost jet region are found in the VLBA data at 43.1\,GHz, 22.2\,GHz, and 15.3\,GHz. 
A single-epoch VLBA observation at 43.1\,GHz produced an image at the highest resolution (at the time of this article) of the jet, revealing a component (unresolved at lower frequencies) very close (0.16\,mas) to the core. 
Stationary components in the vicinity of the radio core are a typical phenomenon in HBLs. 
Combining the SED modeling with this radio behavior, we conclude that this source is a typical HBL even though the synchrotron SED peak lies in the intermediate region when the source is in its lowest state. 

(v) We were able to use a two-component (``blob-in-jet") SSC model to describe multiple flux states of the source. 
The flaring state is sufficiently described with the same model parameters for the jet component as the low state and with a different particle distribution and magnetic field for the blob component. 

(vi) The fluxes measured by the LAT in the HE regime and by Tuorla at optical energies are found to follow a log-normal distribution and to be strongly temporally correlated with one another.  
This is consistent with a SSC emission process.

(vii) We searched for evidence of a periodic signal in the Tuorla optical data and in the {\it{Fermi}}-LAT data, the two datasets for which we have the best-sampled light curves. No evidence for periodicity on any timescale is detected.\\

In the future, studies such as the ones presented should be performed on larger data sets, covering different emission states of the source being studied. Such data are expected to be provided at gamma-ray energies by the Cherenkov Telescope Array \citep{CTA_book2019}. 

\acknowledgments

The {\sl Fermi}-LAT Collaboration acknowledges generous ongoing support from a number of agencies and institutes that have supported both the development and the operation of the LAT as well as scientific data analysis. 
These include the National Aeronautics and Space Administration and the Department of Energy in the United States; the Commissariat \`a l'Energie Atomique and the Centre National de la Recherche Scientifique/Institut National de Physique Nucl\'eaire et de Physique des Particules in France; the Agenzia Spaziale Italiana and the Istituto Nazionale di Fisica Nucleare in Italy; the Ministry of Education, Culture, Sports, Science and Technology (MEXT), High Energy Accelerator Research Organization (KEK), and Japan Aerospace Exploration Agency (JAXA) in Japan; and the KA Wallenberg Foundation, the Swedish Research Council, and the Swedish National Space Board in Sweden. 
Additional support for science analysis during the operations phase is gratefully acknowledged from the Istituto Nazionale di Astrofisica in Italy and the Centre National d'\'Etudes Spatiales in France. This work performed in part under DOE Contract DE- AC02-76SF00515.

This research is supported by grants from the U.S. Department of Energy Office of Science, the U.S. National Science Foundation and the Smithsonian Institution, and by NSERC in Canada. This research used resources provided by the Open Science Grid, which is supported by the National Science Foundation and the U.S. Department of Energy's Office of Science, and resources of the National Energy Research Scientific Computing Center (NERSC), a U.S. Department of Energy Office of Science User Facility operated under Contract No. DE-AC02-05CH11231. We acknowledge the excellent work of the technical support staff at the Fred Lawrence Whipple Observatory and at the collaborating institutions in the construction and operation of VERITAS.

This research has made use of data from the MOJAVE database that is maintained by the MOJAVE team \citep{Lister18}. The MOJAVE program is supported under NASA-{\sl Fermi} grant NNX15AU76G.

The National Radio Astronomy Observatory is a facility of the National Science Foundation operated under cooperative agreement by Associated Universities, Inc.

This research has made use of the NASA/IPAC Extragalactic Database (NED) which is operated by the Jet Propulsion Laboratory, California Institute of Technology, under contract with the National Aeronautics and Space Administration.

This research has made use of data from the OVRO 40-m monitoring program \citep{Richards11} which is supported in part by NASA grants NNX08AW31G, NNX11A043G, and NNX14AQ89G and NSF grants AST-0808050 and AST-1109911.

J.V. was partially supported by the Alliance program, a partnership between Columbia University, NY, USA, and three major French institutions: \'Ecole Polytechnique, Paris 1 Panth\'eon-Sorbonne University and Science Po.

T.S. was supported by the Academy of Finland projects 274477, 284495, and 312496.

W.M. acknowledge support from CONICYT project Basal AFB-170002.

Y.Y.K.\ and A.B.P.\ were supported by the Russian Foundation for Basic Research (project 17-02-00197), the government of the Russian Federation (agreement 05.Y09.21.0018), and the Alexander von Humboldt Foundation.

S.K. acknowledges support from the European Research Council under the
European Union's Horizon 2020 research and innovation program, under
grant agreement No~771282.

\software{AIPS (Greisen 2003), DIFMAP (Shepherd 1997), Fermi Science Tools (Fermi Science Support development team 2019)}

\appendix

\section{``Harder-when-brighter'' trend in the LAT yearly data}\label{app:yearlyfits}

\begin{deluxetable}{l|ccc|ccc}[ht]
\tablecaption{ Results of the fit of the yearly {\it Fermi}-LAT data.\label{tab:index-flux}}
\tabletypesize{\scriptsize}
\tablehead{
\colhead{Model} & \multicolumn{3}{c}{Total} & \multicolumn{3}{c}{Non-flare} \\
\colhead{function} & \colhead{$a$} & \colhead{$b$} & \colhead{$\chi^2$/d.o.f.} & \colhead{$a$} & \colhead{$b$} & \colhead{$\chi^2$/d.o.f.}}
\startdata
Constant  & NA & 1.92$\pm$0.02 & 17.8/7 & NA & 1.93$\pm$0.01 & 14.2/7 \\
Linear & $-(1.61\pm0.35)\times 10^{6}$ & 2.06$\pm$0.03 & 4.5/6 & $-(1.41\pm0.48)\times 10^{6}$ & 2.05$\pm$0.04 & 6.3/6 \\
\hline
Preference  &  &  & $3.6\,\sigma$ &  &  &  $ 2.8\,\sigma$
\enddata
\tablenotetext{}{}
\tablecomments{For a linear function $ax+b$, $a$ is the slope and $b$ is the independent term. For a constant function $a$ is not applicable (NA).}
\end{deluxetable}

Details of the fit of the yearly data, total and outside flares, are found in Table \ref{tab:index-flux}. A weak preference towards a harder-when-brighter trend is observed in both data sets. See discussion in Section \ref{sec:latSpectrumFlux}.

\section{XRT Data log}
\label{app:xrt}

We provide here in Table~\ref{tab:Xspec} a log of the XRT data and results included in this paper.

\begin{deluxetable}{ccccc}
\tablecaption{X-ray spectral analysis. \label{tab:Xspec}}
\tabletypesize{\scriptsize}
\tablehead{
\colhead{Observation} &\colhead{Start date} &\colhead{Energy flux}  & \colhead{Photon index}    & $\chi^{2}/\text{d.o.f.}$                            \\
\colhead{} &\colhead{} &\colhead{($10^{-12}$ erg cm$^{-2}$ s$^{-1}$)}  & \colhead{}   & \colhead{}
}
\startdata
31553001 & 2009 Dec 03 16:18:59 &   $4.94^{+0.17}_{-0.19}$ &  2.52$\,\pm\,$0.06 &  36.2/33.0 \\
 31906001 & 2011 Jan 04 00:49:00 &   $9.96^{+0.39}_{-0.31}$ &  2.37$\,\pm\,$0.04 &  51.6/52.0 \\
 31906002 & 2011 Jan 08 03:04:27 &   $7.85^{+0.38}_{-0.43}$ &  2.19$\,\pm\,$0.07 &  24.1/24.0 \\
 31906004 & 2011 Jan 10 03:11:00 &   $6.37^{+0.26}_{-0.26}$ &  2.45$\,\pm\,$0.06 &  40.8/32.0 \\
 31906005 & 2011 Jan 11 03:16:00 &   $6.74^{+0.23}_{-0.27}$ &  2.50$\,\pm\,$0.05 &  26.7/39.0 \\
 31906006 & 2011 Jan 12 03:20:59 &   $8.36^{+0.36}_{-0.30}$ &  2.32$\,\pm\,$0.06 &  36.3/32.0 \\
 31906007 & 2011 Dec 08 14:11:59 &   $2.77^{+0.24}_{-0.18}$ &  2.36$\,\pm\,$0.14 &    9.8/8.0 \\
 31906008 & 2013 Feb 19 09:47:58 &  $13.88^{+0.75}_{-1.03}$ &  2.17$\,\pm\,$0.09 &  22.9/18.0 \\
 31906009 & 2013 Mar 08 08:59:59 &   $5.86^{+0.54}_{-0.65}$ &  2.66$\,\pm\,$0.17 &    4.7/6.0 \\
 31906010 & 2013 Mar 13 07:34:59 &   $5.48^{+0.53}_{-0.60}$ &  2.47$\,\pm\,$0.17 &    0.6/5.0 \\
 31906011 & 2013 Mar 17 08:06:59 &  $10.81^{+0.75}_{-0.76}$ &  2.37$\,\pm\,$0.10 &   8.9/14.0 \\
 31906012 & 2014 Feb 09 13:31:02 &  $11.61^{+0.33}_{-0.39}$ &  2.37$\,\pm\,$0.05 &  53.4/31.0 \\
 31972001 & 2011 Apr 22 05:27:00 &   $5.79^{+0.40}_{-0.50}$ &  2.67$\,\pm\,$0.13 &    3.9/9.0 \\
 31972002 & 2011 Apr 23 05:51:00 &   $5.11^{+0.28}_{-0.20}$ &  2.65$\,\pm\,$0.10 &   5.9/13.0 \\
 31972003 & 2011 Apr 24 04:30:00 &   $4.80^{+0.26}_{-0.28}$ &  2.76$\,\pm\,$0.09 &  19.7/15.0 \\
 31972004 & 2011 Apr 25 05:56:00 &   $3.76^{+0.20}_{-0.26}$ &  2.62$\,\pm\,$0.12 &   9.5/11.0 \\
 31972006 & 2011 Apr 29 04:23:00 &   $4.04^{+0.25}_{-0.37}$ &  2.57$\,\pm\,$0.13 &  16.3/10.0 \\
 31972007 & 2011 May 01 04:33:00 &   $3.79^{+0.25}_{-0.26}$ &  2.83$\,\pm\,$0.14 &    6.9/9.0 \\
 31972008 & 2011 May 02 04:37:59 &   $3.46^{+0.37}_{-0.64}$ &  2.39$\,\pm\,$0.20 &    7.0/5.0 \\
 31972010 & 2011 May 05 03:40:00 &   $4.54^{+0.38}_{-0.27}$ &  2.80$\,\pm\,$0.13 &  11.5/10.0 \\
 31972002 & 2011 Apr 23 05:51:00 &   $6.16^{+0.90}_{-1.08}$ &  2.75$\,\pm\,$0.32 &   20.5/16.0 \\
 31906013 & 2017 Apr 15 11:33:21 &  $28.55^{+0.61}_{-0.66}$ &  2.72$\,\pm\,$0.04 &  129.0/95.0 \\
 31906014 & 2017 Apr 17 17:12:06 &  $33.08^{+0.79}_{-0.76}$ &  2.64$\,\pm\,$0.03 &  101.1/98.0 \\
 31906015 & 2017 Apr 19 02:35:57 &  $24.63^{+1.03}_{-0.62}$ &  3.02$\,\pm\,$0.06 &   83.6/52.0 \\
 31906016 & 2017 Apr 23 13:24:57 &  $32.30^{+1.17}_{-1.07}$ &  2.51$\,\pm\,$0.04 &   74.6/66.0 \\
\enddata
\end{deluxetable}

\section{Long-term {\sl Fermi}-LAT SED Data}
\label{app:LATsed9y}

Details of the LAT long-term spectral analysis results are provided in Table \ref{tab:LATsed9y}. These data are shown in Figure \ref{fig:9yrsLATsed} in Section \ref{sec:longtermsed}.

\begin{deluxetable*}{cccc}[ht]
\tablecaption{{\sl Fermi}-LAT spectral analysis. Time range: 2008 August 04\,-\,2017 September 04. \label{tab:LATsed9y}}
\tabletypesize{\scriptsize}
\tablehead{
\colhead{Energy range} & Significance & Integral flux & Predicted counts      \\
\colhead{(GeV)} &\colhead{} &\colhead{(ph cm$^{-2}$ s$^{-1}$)}  & \colhead{}
}
\startdata
0.100 -	0.146	&	10.92	&	(1.62 $\pm$ 0.48)$\times 10^{-8}$ &	1893.8	 \\
0.146 -	0.215	&	22.06	&	(1.65 $\pm$ 0.19)$\times 10^{-8}$ &	2766.9  \\
0.215 -	0.316	&	24.84	&	(1.03 $\pm$ 0.09)$\times 10^{-8}$ &	2198.6	 \\
0.316 -	0.464	&	30.74	&	(7.36 $\pm$ 0.44)$\times 10^{-9}$ &	1857.2	 \\
0.464 -	0.681	&	38.58	&	(5.58 $\pm$ 0.25)$\times 10^{-9}$ &	 1608.2 \\
0.681 -	1		&	44.91	&	(4.12 $\pm$ 0.16)$\times 10^{-9}$ &	 1319.3	\\ 
1 -	1.467		&	47.29	&	(2.81 $\pm$ 0.11)$\times 10^{-9}$ &	 992.1	 \\
1.467 -	2.154	&	51.77	&	(2.20 $\pm$ 0.09)$\times 10^{-9}$ &	 820.0	 \\
2.154 -	3.162	&	50.36	&	(1.48 $\pm$ 0.07)$\times 10^{-9}$ &	 559.7	 \\
3.162 -	4.641	&	43.85	&	(9.82 $\pm$ 0.55)$\times 10^{-10}$ & 366.0  \\
4.641 -	6.812	&	39.99	&	(7.07 $\pm$ 0.47)$\times 10^{-10}$ & 258.5	 \\
6.812 -	10		&	32.37	&	(4.23 $\pm$ 0.36)$\times 10^{-10}$ & 155.8	 \\
10 -	14.677	&	32.49	&	(3.52 $\pm$ 0.32)$\times 10^{-10}$ & 129.9  \\
14.677 - 21.544	&	27.86	&	(2.50 $\pm$ 0.27)$\times 10^{-10}$ & 91.4	 \\
21.544 - 31.622	&	20.40	&	(1.22 $\pm$ 0.18)$\times 10^{-10}$ & 45.2	 \\
31.622 - 46.415	&	16.17	&	(8.02 $\pm$ 1.50)$\times 10^{-11}$ & 30.5	 \\
46.415 - 68.129	&	18.43	&	(8.51 $\pm$ 1.54)$\times 10^{-11}$ & 32.4	 \\
68.129 - 100	&	12.87	&	(4.10 $\pm$ 1.09)$\times 10^{-11}$ & 15.6	 \\
100 -	500		&	7.33	&	(2.19 $\pm$ 0.79)$\times 10^{-11}$ & 8.2	 
\enddata
\end{deluxetable*}

\bibliography{QiBibAll}
\end{document}